\documentclass[12pt]{article}
\usepackage[latin1]{inputenc}
\usepackage[dvips]{graphicx,color}
\usepackage{amssymb}
\usepackage{hyperref}

\def\beq{\begin{equation}}
\def\eeq{\end{equation}}
\def\eq{\end{equation}}
\def\ba{\begin{eqnarray}}
\def\ea{\end{eqnarray}}

\def\centeron#1#2{{\setbox0=\hbox{#1}\setbox1=\hbox{#2}\ifdim
\wd1>\wd0\kern.5\wd1\kern-.5\wd0\fi
\copy0\kern-.5\wd0\kern-.5\wd1\copy1\ifdim\wd0>\wd1
\kern.5\wd0\kern-.5\wd1\fi}}
\def\ltap{\;\centeron{\raise.35ex\hbox{$<$}}{\lower.65ex\hbox{$\sim$}}\;}
\def\gtap{\;\centeron{\raise.35ex\hbox{$>$}}{\lower.65ex\hbox{$\sim$}}\;}

\newcommand{\captionfonts}{\small}
\makeatletter \long\def\@makecaption#1#2{
  \vskip\abovecaptionskip
  \sbox\@tempboxa{{\captionfonts #1: #2}}
  \ifdim \wd\@tempboxa >\hsize
    {\captionfonts #1: #2\par}
  \else
    \hbox to\hsize{\hfil\box\@tempboxa\hfil}
    \fi
  \vskip\belowcaptionskip}
\makeatother

\newcommand{\newc}{\newcommand}
\newc{\qbar}{{\overline q}}
\newc{\Kahler}{K\"ahler }
\newc{\deltaGS}{\delta_{\rm GS}}
\newc{\bsg}{B\rightarrow X_s\gamma}
\newc{\Bmumu}{B_s\rightarrow \mu^+ \mu^-}
\newc{\MSusy}{m_{\tilde{t}}}
\newc{\HEWSB}{H^{\textrm{\tiny{EWSB}}}}
\newc{\cossqbma}{\cos^2(\beta-\alpha)}
\newc{\sinsqbma}{\sin^2(\beta-\alpha)}
\newc{\HSM}{H_{\textrm{\tiny{SM}}}}
\newc{\MSusyMin}{m_{\tilde{t},\textrm{\tiny{min}}}}

\begin{document}
\def\pplogo{\vbox{\kern-\headheight\kern -29pt
\halign{##&##\hfil\cr&{\ppnumber}\cr\rule{0pt}{2.5ex}&\ppdate\cr}}}
\makeatletter
\def\ps@firstpage{\ps@empty \def\@oddhead{\hss\pplogo}
  \let\@evenhead\@oddhead
}
\def\maketitle{\par
 \begingroup
 \def\thefootnote{\fnsymbol{footnote}}
 \def\@makefnmark{\hbox{$^{\@thefnmark}$\hss}}
 \if@twocolumn
 \twocolumn[\@maketitle]
 \else \newpage
 \global\@topnum\z@ \@maketitle \fi\thispagestyle{firstpage}\@thanks
 \endgroup
 \setcounter{footnote}{0}
 \let\maketitle\relax
 \let\@maketitle\relax
 \gdef\@thanks{}\gdef\@author{}\gdef\@title{}\let\thanks\relax}
\makeatother


\setcounter{page}0
\def\ppnumber{\vbox{\baselineskip14pt
}}
\def\ppdate{RUNHETC-06-18} \date{}

\author{\\Rouven Essig\footnote{rouven@physics.rutgers.edu}\\
[7mm]
{\normalsize NHETC, Department of Physics and Astronomy,}\\
{\normalsize Rutgers University, Piscataway, NJ 08854, U.S.A.}\\}

\title{\bf \LARGE Implications of the LEP
Higgs Bounds \\ for the MSSM Stop Sector} \maketitle \vskip 2cm

\begin{abstract} \normalsize
\noindent  The implications of the LEP Higgs bounds on the MSSM stop
masses and mixing are compared in two different regions of the Higgs
parameter space. The first region is the Higgs decoupling limit, in
which the bound on the mass of the lighter Higgs is $m_h \ge 114.4$
GeV, and the second region is near a non-decoupling limit with $m_h
\simeq 93$ GeV, in which the masses of all the physical Higgs bosons
are required to be light. Additional constraints from the
electroweak $S$- and $T$-parameter and the decays $B \to X_s \gamma$
and $B_s \to \mu^+ \mu^-$, which also constrain the Higgs and/or
stop sector, are considered.  In some regions of the MSSM parameter
space these additional constraints are stronger than the LEP Higgs
bounds. Implications for the tuning of electroweak symmetry breaking
are also discussed.
\end{abstract}
\bigskip

\newpage

\tableofcontents


\section{Introduction}\label{Sec:Intro}

The Higgs sector in the Minimal Supersymmetric Standard Model (MSSM)
consists of two $SU(2)_L$ doublets, $H_d$ and $H_u$, with opposite
hypercharge. Five physical states remain after electroweak symmetry
breaking (EWSB).  Assuming there are no CP-violating phases, these
physical states consist of two neutral CP-even states $h$ and $H$
with masses $m_h \le m_H$, one neutral CP-odd state $A$, and two
charged states $H^{\pm}$.  The tree-level masses of $h$ and $H$ are
bounded, $m^{\textrm{\tiny{tree}}}_h \le m_Z \le
m^{\textrm{\tiny{tree}}}_H$, with $m_Z\simeq 91$ GeV.

Using the Large Electron Positron (LEP) collider, the LEP
collaboration searched for these Higgs bosons and published bounds
on their masses \cite{ALEPH:2006cr}. Their results have ruled out
substantial regions of the MSSM Higgs parameter space, and in much
of the remaining allowed regions it is clear that large radiative
corrections to the tree-level Higgs masses are required to satisfy
the LEP bounds. However, two very different scenarios are still
possible. One scenario is obtained in the Higgs decoupling limit in
which $h$ behaves like the Standard Model (SM) Higgs and all the
other Higgs bosons become heavy and decouple from the low energy
theory. Here the bound on $m_h$ coincides with the bound on the mass
of the SM Higgs, namely $m_h \ge 114.4$~GeV \cite{Barate:2003sz}.
The other scenario is obtained in the Higgs ``non-decoupling'' limit
in which $H$ behaves like the SM Higgs and the Higgs sector is
required to be light.  It allows for 93 GeV $\lesssim m_h <$ 114.4
GeV, where the value of 93 GeV is the (somewhat model dependent)
lower bound that the LEP collaboration has obtained for $m_h$
assuming various decay scenarios for $h$ and a variety of different
``benchmark'' parameter choices for the MSSM parameters. If $m_h$ is
near 93 GeV, it seems to naively require much smaller radiative
corrections to the tree-level Higgs mass than when $m_h$ is near
114.4~GeV. Since the dominant radiative corrections to the
tree-level CP-even Higgs mass matrix, which determines $m_h$ and
$m_H$, come from loops involving the top quark and stop squarks, one
might naively suspect that $m_h$ near 93 GeV allows for much smaller
stop masses than $m_h$ near 114.4 GeV. Moreover, larger stop masses
would in general imply a more fine-tuned MSSM, and one might
therefore suspect that the MSSM is less fine-tuned for $m_h$ near 93
GeV.  In this paper, we present lower bounds on the stop masses
consistent with the LEP Higgs bounds, both in the Higgs decoupling
region, with $m_h \ge 114.4$ GeV, as well as near the Higgs
non-decoupling region, with $m_h \simeq 93$ GeV. We compare the
constraints on the stop masses in these two regions of the Higgs
parameter space, and show that in certain regions of the MSSM
parameter space the lower bounds on the stop masses are not
significantly different from each other. Furthermore, although there
are regions in which the lower bounds are smaller, there are also
regions in which they are larger.

There are other constraints on new physics that may further tighten
bounds on the stop or Higgs sector. These additional constraints
include the electroweak $S$- and $T$-parameter, and the decays
$\bsg$ and $\Bmumu$. In this paper, we discuss the regions of the
MSSM parameter space in which these additional constraints are
important in restricting the stop and/or Higgs sector further.

The outline of this paper is as follows.  In Section \ref{Sec:LEP
and MSSM Higgs sector}, we investigate the LEP constraints on the
neutral MSSM Higgs sector and its implication for the stop sector in
more detail.  This will allow us to obtain a simple numerical
estimate of the lower bound on the stop masses in a particular limit
of the MSSM parameter space.  This estimate is independent of the
size of $m_h$. Section \ref{Sec:Lower bounds on MSusy} contains the
main results of this paper.  We give lower bounds on the stop masses
consistent with the LEP Higgs bounds.  The analysis will include all
the important radiative corrections to the CP-even Higgs mass
matrix, and we discuss the importance of the top mass, the stop
mixing the gaugino masses and the supersymmetric Higgsino mass
parameter ($\mu$) on the lower bounds of the stop masses. In
addition, we present results on how the lower bounds on the soft
stop masses vary in the decoupling limit as a function of $m_h$. In
Section \ref{Sec:Constraints from new physics}, we discuss how other
constraints on new physics impact the results in Section
\ref{Sec:Lower bounds on MSusy}.  In particular, we investigate the
effect of the electroweak $S$- and $T$-parameters, $\bsg$, and
$\Bmumu$.  Section \ref{Sec:Implications for EWSB} contains a
discussion of the implications of our analysis for electroweak
symmetry breaking and the supersymmetric little hierarchy problem.
In Section \ref{Sec:Conclusion}, we summarize the results of this
paper. Appendix \ref{Sec:MSSM Higgs sector and LEP} gives the
relevant background to understand the LEP results for the MSSM Higgs
sector. In Appendix \ref{Sec:At fixed point}, we review the
quasi-fixed point for the stop soft trilinear coupling, $A_t$.  The
trilinear coupling is the main ingredient in determining the amount
of stop mixing, and we use the quasi-fixed point value for $A_t$ in
some of the main results of this paper.

\section{LEP constraints on the Higgs sector and implications for the MSSM stop sector}\label{Sec:LEP and MSSM Higgs
sector}

In this section, we first review the LEP Higgs constraints, before
going on to discuss the implications of these constraints for the
stop sector.

\subsection{Constraints from LEP on the MSSM Higgs-sector}

The LEP collaboration searched for the production of Higgs bosons in
both the Higgsstrahlung ($e^+e^-$ $\rightarrow$ $Z$ $\rightarrow$
$Zh$ (or $ZH$)) and pair production ($e^+e^-$ $\rightarrow$ $Z$
$\rightarrow$ $Ah$ (or $AH$)) channels.  The results from these
channels have been used to set upper bounds on the couplings $ZZh$
($ZZH$) and $ZAh$ ($ZAH$) as a function of the Higgs masses. These
couplings are proportional to either $\sin^2(\beta-\alpha)$ or
$\cos^2(\beta-\alpha)$.  Here $\beta$ is determined from the ratio
of the two vacuum expectation values $v_u\equiv\langle{\rm Re}
(H_u^0)\rangle$ and $v_d\equiv\langle{\rm Re} (H_d^0)\rangle$ as
$\tan\beta = v_u/v_d$, and $\alpha$ is the neutral CP-even Higgs
mixing angle.

Within the MSSM, the results from the Higgsstrahlung channel give an
upper bound on $\sin^2(\beta-\alpha)$ and $\cossqbma$ as a function
of $m_h$ and $m_H$, respectively (see Fig.~2 in
\cite{ALEPH:2006cr}).  The pair production channel, on the other
hand, gives an upper bound on $\cos^2(\beta-\alpha)$ and $\sinsqbma$
as a function of $m_h + m_A$ and $m_H + m_A$, respectively (see
Fig.~4 in \cite{ALEPH:2006cr}). Appendix \ref{Sec:MSSM Higgs sector
and LEP} contains a review on how these functions of $\alpha$ and
$\beta$ appear in the MSSM, and why LEP bounds them.

The LEP results from the Higgsstrahlung channel put several
interesting bounds on $m_h$, $m_H$ and $\sinsqbma$. In the
decoupling limit, $h$ behaves like the SM Higgs so that $\sinsqbma
\rightarrow 1$ and the bound on its mass is given by

\beq\label{Eqn:lower Region A} m_h \ge 114.4 {\textrm{
GeV}},~\sinsqbma \rightarrow 1. \eeq

If $m_h$ is less than 114.4 GeV, smaller values of $\sinsqbma$ are
required in order to suppress the production of $h$ in the
Higgsstrahlung channel and to allow it to have escaped detection.
For $m_h$ $\simeq$ 93 GeV, $\sinsqbma$ needs to be less than about
0.2, so that $\cossqbma \gtrsim 0.8$ (from Fig.~2 in
\cite{ALEPH:2006cr}). Larger values of $\cossqbma$, however,
increase the $HZZ$ coupling so that now $m_H$ needs to be large
enough to suppress the production of $H$ in the Higgsstrahlung
channel, and allow it, in turn, to have escaped detection.  We find
that $m_H \gtrsim 114.0$ GeV (from Fig.~2 in \cite{ALEPH:2006cr}).
If $\sinsqbma$ is even smaller and approaches zero, i.e.~$\cossqbma
\rightarrow 1$, it is $H$ which behaves like the SM Higgs so that
the bound on its mass is given by $m_H$ $\ge$ 114.4 GeV (this will
be referred to as the Higgs ``non-decoupling'' limit).

In Section \ref{Sec:Lower bounds on MSusy}, we present lower bounds
on the stop soft masses for two regions in the Higgs parameter
space. These two regions are given by equation (\ref{Eqn:lower
Region A}) and by

\begin{equation} \label{Eqn:lower Region B}
m_h \simeq 93 {\textrm{ GeV}}, ~\cossqbma \ge 0.8, ~m_H \ge 114.4
\textrm{ GeV}.
\end{equation}

\noindent (We choose $m_H$ in equation (\ref{Eqn:lower Region B}) to
be at least above 114.4 GeV, in order
to allow the full range $0.8 \le \cossqbma \le 1$.)

We note that the bounds given in the previous paragraphs assume that
the MSSM Higgs boson $h$ decays like the SM Higgs boson (see
\cite{ALEPH:2006cr}). If we assume different Higgs decay branching
ratios, somewhat different bounds can be obtained. For example,
assuming $h$ decays completely into $\tau\bar{\tau}$ gives a
stricter bound on the $hZZ$ coupling for a wide range of $m_h$.  The
LEP collaboration even considered the extreme case in which the
Higgs decays invisibly. In this case, the bound on the $hZZ$
coupling as a function of $m_h$ is in general not much worse than if
we assume that the Higgs decays like a SM Higgs. In fact, for some
range of $m_h$ the bound is even stricter if we assume that the
Higgs decays invisibly \cite{LHWG:2001xz}.

The lower bound on $m_h$ is also model dependent.  For example, in
\cite{ALEPH:2006cr}, figures are presented that show excluded
regions in the MSSM parameter space for a variety of ``benchmark''
scenarios that consist of different choices for the MSSM parameters.
The LEP collaboration found that the lower bound on $m_h$ can be
slightly less than 93 GeV in certain cases. Moreover, the authors in
\cite{Belyaev:2006rf} claim that there are certain regions in
parameter space for which the $ZZh$ coupling and the $h/A\rightarrow
b\bar{b}$ branching ratios are both suppressed and that this allows
$m_h$ to be substantially less than 93 GeV.

\subsection{Implications for the MSSM stop sector}\label{Sec:Implication for MSSM stop sector}

Since the tree-level mass of the lighter neutral Higgs is bounded
above by $m_Z$, it is clear that substantial radiative corrections
are required to push the lighter Higgs mass above 114.4 GeV in the
Higgs decoupling limit.

We now discuss why substantial radiative corrections to the
tree-level Higgs masses are also required if the lighter Higgs mass
is near 93 GeV.  This is most easily seen in the large $\tan\beta$
limit. In this limit, the CP-even Higgs mass squared matrix in the
($H_d$,$H_u$) basis is particularly simple if we only include the
tree-level pieces and the dominant radiative corrections.  Since
$\tan\beta$ is large, the vacuum expectation value $v_d$ vanishes in
this limit, and the Higgs vacuum expectation value is thus
completely determined by $v_u$.  In the absence of any radiative
corrections, one of the physical Higgs mass eigenstates lies
completely in the $H_u$ direction and thus behaves like the SM Higgs
(with a mass equal to $m_Z$), whereas the other mass eigenstate lies
completely in the $H_d$ direction (with a mass equal to the mass of
the CP-odd Higgs, $m_A$). This alignment of the two physical CP-even
Higgs mass eigenstates with the $H_u$ and $H_d$ direction,
respectively, remains unchanged when only the dominant radiative
correction is added. The reason for this is that, due to the large
top Yukawa coupling, the dominant radiative corrections to the Higgs
sector are to the up-type Higgs soft supersymmetry breaking
Lagrangian mass and come from loops involving the top quark and stop
squarks \cite{Okada:1990vk,Ellis:1990nz,Haber:1990aw}.  This gives a
correction to the $H_u$-$H_u$ component of the CP-even Higgs mass
squared matrix. The matrix is thus particularly simple for large
$\tan\beta$, and is given by

\begin{equation}
\mathcal{M}^2 \approx \left( \begin{array}{cc} m_A^2 & 0 \\
0 & m_Z^2 + \delta \mathcal{M}^2_{uu}
\end{array} \right) ~~~~~~~({\textrm{for large}}~ \tan\beta),
\end{equation}

\noindent where $\delta \mathcal{M}^2_{uu}$ is the dominant top/stop
correction.  Since the $H_u$-$H_u$ component for large $\tan\beta$
gives the mass of the physical Higgs that behaves like the SM Higgs,
its value is bounded below by 114.4 GeV, i.e.

\begin{equation}\label{Eqn:Size of delta M2uu}
m_Z^2 + \delta \mathcal{M}^2_{uu} \gtrsim (114.4 \textrm{ GeV})^2.
\end{equation}

\noindent The result in equation (\ref{Eqn:Size of delta M2uu}) is
independent of whether the lighter or the heavier Higgs lies in the
$H_u$ direction (this depends on the size of $m_A$). It also shows
that the lower bound on the size of the required radiative
corrections is fixed and independent of the mass of the lighter
Higgs, at least in the large $\tan\beta$ limit including only the
leading corrections.  Moreover,  it is $m_H$ which acquires the
dominant radiative corrections for $m_h\simeq$ 93 GeV.

A simple estimate of the lower bounds on the stop masses in the
large $\tan\beta$ limit may be obtained using equation
(\ref{Eqn:Size of delta M2uu}).  For large $\tan\beta$, the dominant
radiative correction is given by

\begin{equation}\label{Eqn:delta_M_uu}
\delta\mathcal{M}^2_{uu} \simeq \frac{3 g^2 m_t^4}{8\pi^2 m_W^2}
\Bigg\{\ln\Bigg(\frac{\MSusy^2}{m_t^2}\Bigg) +
\frac{X_t^2}{\MSusy^2}\Bigg(1-\frac{X_t^2}{12
\MSusy^2}\Bigg)\Bigg\},
\end{equation}

\noindent where $m_t$ is the top mass, $g$ is the $SU(2)_L$ gauge
coupling, and $m_W$ is the mass of the $W$-bosons
\cite{Okada:1990vk,Ellis:1990nz,Haber:1990aw}. Furthermore, equation
(\ref{Eqn:delta_M_uu}) assumes that the stop soft masses are equal
to $\MSusy$, with $\MSusy$ $\gg$ $m_t$. The stop mixing parameter is
given by $X_t=A_t - \mu\cot\beta$ ($\simeq$ $A_t$ for large
$\tan\beta$), where $A_t$ denotes the stop soft trilinear coupling
and $\mu$ is the supersymmetric Higgsino mass parameter. The
dependence on the top mass to the fourth power is particularly
noteworthy.  The first term in equation (\ref{Eqn:delta_M_uu}) comes
from renormalization group running of the Higgs quartic coupling
below the stop mass scale and vanishes in the limit of exact
supersymmetry. It grows logarithmically with the stop mass. The
second term is only present for non-zero stop mixing and comes from
a finite threshold correction to the Higgs quartic coupling at the
stop mass scale.  It is independent of the stop mass for fixed
$X_t/\MSusy$, and grows linearly as $(X_t/\MSusy)^2$ for small
$X_t/\MSusy$.

It is apparent from equation (\ref{Eqn:delta_M_uu}) that the mixing
term is important for determining lower bounds on the stop masses.
Using equation (\ref{Eqn:delta_M_uu}) and assuming no mixing
($X_t=0$), we require $\MSusy \gtrsim 570$ GeV in order to satisfy
the LEP bound in (\ref{Eqn:Size of delta M2uu}). This value was
obtained using a running top mass of $m_t(m_t) \simeq 167$ GeV
\cite{Haber:1996fp}. The second (mixing) term in equation
(\ref{Eqn:delta_M_uu}), however, reaches a maximum of 3 for $X_t =
\sqrt{6}\MSusy$, called \emph{maximal-mixing}. In order for the
logarithm of the first term to be of the same order, $\MSusy$ needs
to be about 750 GeV.  Thus the mixing term alone is more than enough
to give the required radiative corrections to satisfy the LEP bound.
Mixing in the stop sector therefore allows for much smaller stop
masses.

There are other radiative corrections to the Higgs masses which are
important, including negative radiative corrections that come from
charginos, for example. In Section \ref{Sec:Lower bounds on MSusy},
we include all the important radiative corrections to determine more
accurate lower bounds on the stop masses.  For example, for no stop
mixing and $\tan\beta = 50$, a more accurate lower bound is given by
$\MSusy \gtrsim$ 980 GeV, assuming a physical top mass of 173 GeV,
$\mu = 200$ GeV, and a bino and wino mass of 100 GeV and 200 GeV,
respectively.  This shows the importance of including higher order
corrections to the Higgs sector.  Moreover, the lower bound is
approximately the same for $m_h\simeq 93$ GeV and for $m_h\ge 114.4$
GeV, as expected for large $\tan\beta$.

The above discussion assumes that $\tan\beta$ is large. In Section
\ref{Sec:Lower bounds on MSusy}, we obtain lower bounds on $\MSusy$
also for small and moderate values of $\tan\beta$, for which the
off-diagonal elements in the Higgs mass matrix become important. In
general, we find that the stop masses and/or mixing have to be
sizeable for all values of $\tan\beta$ and for both the Higgs
decoupling and non-decoupling regions. However, depending on the
size of the stop mixing, the lower bounds on the stop masses for
moderate values of $\tan\beta$ can be smaller for $m_h \simeq 93$
GeV than for $m_h \ge 114.4$ GeV (see also \cite{Kim:2006mb}).
Moreover, for small values of $\tan\beta$, the lower bounds on the
stop masses become larger for $m_h \simeq 93$~GeV than for $m_h \ge
114.4$~GeV.

\section{Lower bounds on the stop masses}\label{Sec:Lower bounds on MSusy}

In this section, we present lower bounds on the stop masses
consistent with the LEP Higgs bounds, and we discuss their
dependence on some of the other MSSM parameters. In particular, we
set lower bounds on the left-handed and right-handed stop soft mass,
$m_{\tilde{t}_L}$ and $m_{\tilde{t}_R}$, respectively, taking both
equal to a common value, which we denote by $m_{\tilde{t}}$.  We
denote the lower bound on $\MSusy$ consistent with the LEP Higgs
bounds by $\MSusyMin$.  We consider the two scenarios given in
equations (\ref{Eqn:lower Region A}) and (\ref{Eqn:lower Region B}),
namely the Higgs decoupling limit with $m_h \ge 114.4$ GeV (Section
\ref{Sec:m_h_greater_114_GeV}), and near the Higgs non-decoupling
limit with $m_h\simeq 93$ GeV, and the additional constraints
$\cossqbma\ge 0.8$ and $m_H \ge 114.4$ GeV (Section
\ref{Sec:m_h_simeq_93_GeV}).  In addition, in Section
\ref{Sec:various mh}, we give lower bounds on the stop soft masses
as a function of the physical Higgs boson mass $m_h$ in the
decoupling limit.  All the lower bounds on the stop masses that we
present are consistent with the 2$\sigma$ constraint on $\delta\rho$
(which is related to the electroweak $T$-parameter). In Section
\ref{Sec:Constraints from new physics}, we discuss the importance of
this parameter, as well as others, in constraining the stop masses.

\subsection{Lower bounds on the stop masses for $m_h \ge 114.4$ GeV} \label{Sec:m_h_greater_114_GeV}

For a given set of parameters, we minimize $\MSusy$ by starting it
at the lowest value that gives physical stop masses above 100 GeV
and increasing it until $m_h$ is above 114.4 GeV. We choose the
physical stop mass to be at least 100 GeV since this bound is
illustrative of the actual, slightly model dependent, lower bound
obtained from the Tevatron \cite{PDBook}. The Higgs masses were
calculated with version 2.2.7 of the program
\href{http://www.feynhiggs.de}{$\mathtt{FeynHiggs}$} which includes
all the important radiative corrections to the Higgs sector
\cite{Heinemeyer:1998yj,Heinemeyer:1998np,Degrassi:2002fi,Frank:2006yh}.
We set $m_A = 1000$ GeV to ensure that we are in the Higgs
decoupling limit.

\begin{figure}\begin{center}\includegraphics[scale=0.47]{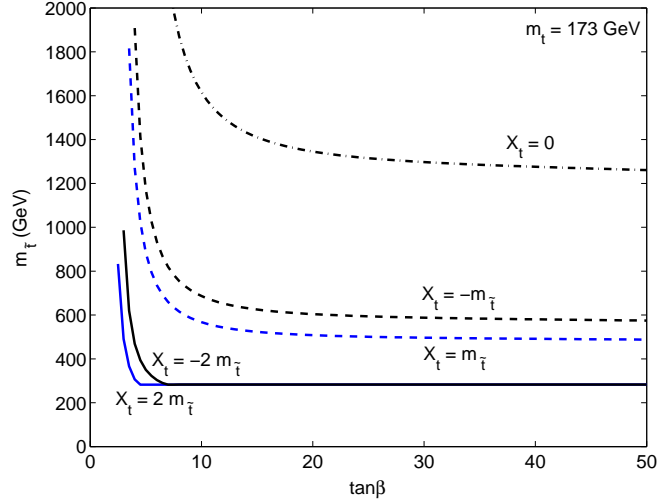}
\caption{Minimum stop soft masses, $m_{\tilde{t}} \equiv
m_{\tilde{t}_L} = m_{\tilde{t}_R} $, for $m_h \geq 114.4$ GeV as a
function of $\tan \beta$ for stop mixing $X_t/ m_{\tilde{t}} = 0,\pm
1, \pm 2$. All squark, slepton, and gaugino soft masses are equal to
the stop soft masses, $\mu=200$ GeV, $m_A = 1000$ GeV, $m_t = 173$
GeV, and all soft trilinear couplings are equal to
$A_t=X_t+\mu\cot\beta$.
\label{Fig:M_S_vs_TB_114_mtop173_all_mixings}}
\end{center}\end{figure}

In Fig.~\ref{Fig:M_S_vs_TB_114_mtop173_all_mixings}, we show
$\MSusyMin$ as a function of $\tan\beta$ for stop mixing $X_t/
m_{\tilde{t}}$ $=$ $0$, $\pm 1$, and $\pm 2$. All squark, slepton,
and gaugino soft masses are equal to $\MSusy$, $\mu=200$ GeV, $m_t =
173$ GeV, and all the soft trilinear couplings are equal to the stop
soft trilinear coupling, $A_t=X_t+\mu\cot\beta$. The lower solid
line shows the \emph{maximal-mixing} scenario,\footnote{The word
``maximal'' refers to the size of the radiative corrections, not to
the amount of mixing. Maximal mixing in $\mathtt{FeynHiggs}$ is
obtained by setting $X_t \simeq 2 \MSusy$, and not $X_t = \sqrt{6}
\MSusy$ as in Section \ref{Sec:Implication for MSSM stop sector}. In
the former case, $X_t$ is defined in the on-shell scheme used in the
diagrammatic two-loop results incorporated into
$\mathtt{FeynHiggs}$, whereas in the latter it is defined in the
$\overline{MS}$-scheme used in the RG approach. Moreover, $m_h$ is
not symmetric with respect to $X_t$ in the full two-loop
diagrammatic calculation in the on-shell scheme. For example, $m_h$
can be up to 5 GeV larger for $X_t=+2 \MSusy$ than for $X_t = -2
\MSusy$. The difference arises from non-logarithmic two-loop
contributions to $m_h$, see \cite{Carena:2000dp,Heinemeyer:2004ms}.}
$X_t = 2 \MSusy$, which approximately maximizes the radiative
corrections to the Higgs sector for a given set of parameters
\cite{Heinemeyer:1999be}. The dot-dashed line shows the
\emph{no-mixing} scenario, $X_t=0$, which approximately minimizes
the radiative corrections to the Higgs sector for a given set of
parameters.  The lower dashed line shows the results for
$X_t=\MSusy$. An \emph{intermediate-mixing} scenario with
$X_t=-\MSusy$ is represented by the upper dashed line. We choose
this scenario since $A_t$ has a strongly attractive infrared
quasi-fixed point at $A_t$ = $-M_3$, see Appendix \ref{Sec:At fixed
point}.  Thus, $A_t$ prefers to be negative due to renormalization
group evolution from the high scale down to the low scale (we choose
the convention in which $M_3$ is positive). In addition, we consider
a scenario which maximizes the Higgs mass for negative stop mixing,
and call it \emph{natural maximal mixing}. This scenario is given by
$X_t = -2 \MSusy$ and is represented by the upper solid-line in the
figure.

A feature that is common to all the curves is that $\MSusy$ becomes
very large for small $\tan\beta$. This is because the tree-level
contribution to the Higgs mass in the decoupling limit is given by
$m_h^{\textrm{\tiny{tree}}}$ $\simeq$ $|\!\cos 2\beta|m_Z$, and goes
to zero as $\tan\beta$ approaches 1. Larger radiative corrections,
and thus larger stop masses, are therefore required for smaller
$\tan\beta$ to push $m_h$ above 114.4 GeV.

Fig.~\ref{Fig:M_S_vs_TB_114_mtop173_all_mixings} clearly shows that
mixing in the stop sector has a large impact on the values of
$\MSusyMin$, with larger mixing allowing much smaller values of
$\MSusyMin$ (see also \cite{Kitano:2006gv,Dermisek:2007fi}). For
large $\tan\beta$, the difference in $\MSusyMin$ between no mixing
and maximal mixing is about 1000 GeV, with $\MSusyMin=1260$ GeV for
$\tan\beta=50$ in the no-mixing case.

\begin{figure}\begin{center}\includegraphics[scale=0.47]{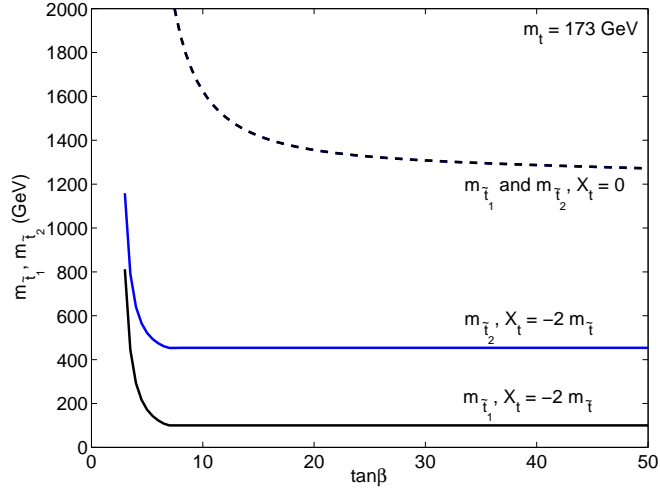}
\caption{Minimum physical stop masses, $m_{\tilde{t}_1}$ and
$m_{\tilde{t}_2}$, for $m_h \geq 114.4$ GeV as a function of $\tan
\beta$ for vanishing stop mixing ($X_t = 0$) and natural maximal
stop mixing ($X_t/ m_{\tilde{t}}= -2$). Other parameters are as
given in Fig.~\ref{Fig:M_S_vs_TB_114_mtop173_all_mixings}.
\label{Fig:stopmasses_vs_TB_mtop173_various_mixings}}
\end{center}\end{figure}

A plot of the two \emph{physical} stop masses, $m_{\tilde{t}_1}$ and
$m_{\tilde{t}_2}$, versus $\tan\beta$ is given in
Fig.~\ref{Fig:stopmasses_vs_TB_mtop173_various_mixings} for no
mixing and for natural maximal mixing.  For no mixing, there is no
discernible difference in the two stop masses since the only
difference that arises is from small $SU(2)_L$ and $U(1)_Y$ D-term
quartic interactions.  For appreciable mixing, the two physical stop
masses are split by an amount that is on the order of $\sim\sqrt{m_t
X_t}$. For $X_t = -2 \MSusy$ and $\tan\beta \ge 7$, $\MSusyMin$ is
small enough that the lighter physical stop mass is all the way down
at its experimental lower bound of roughly 100 GeV. For this range
of $\tan\beta$, we find that $\MSusy$ is larger than that which is
required to get $m_h$ just above 114.4 GeV, and thus $m_h$ is
several GeV above 114.4 GeV here.

\begin{figure}\begin{center}\includegraphics[scale=0.47]{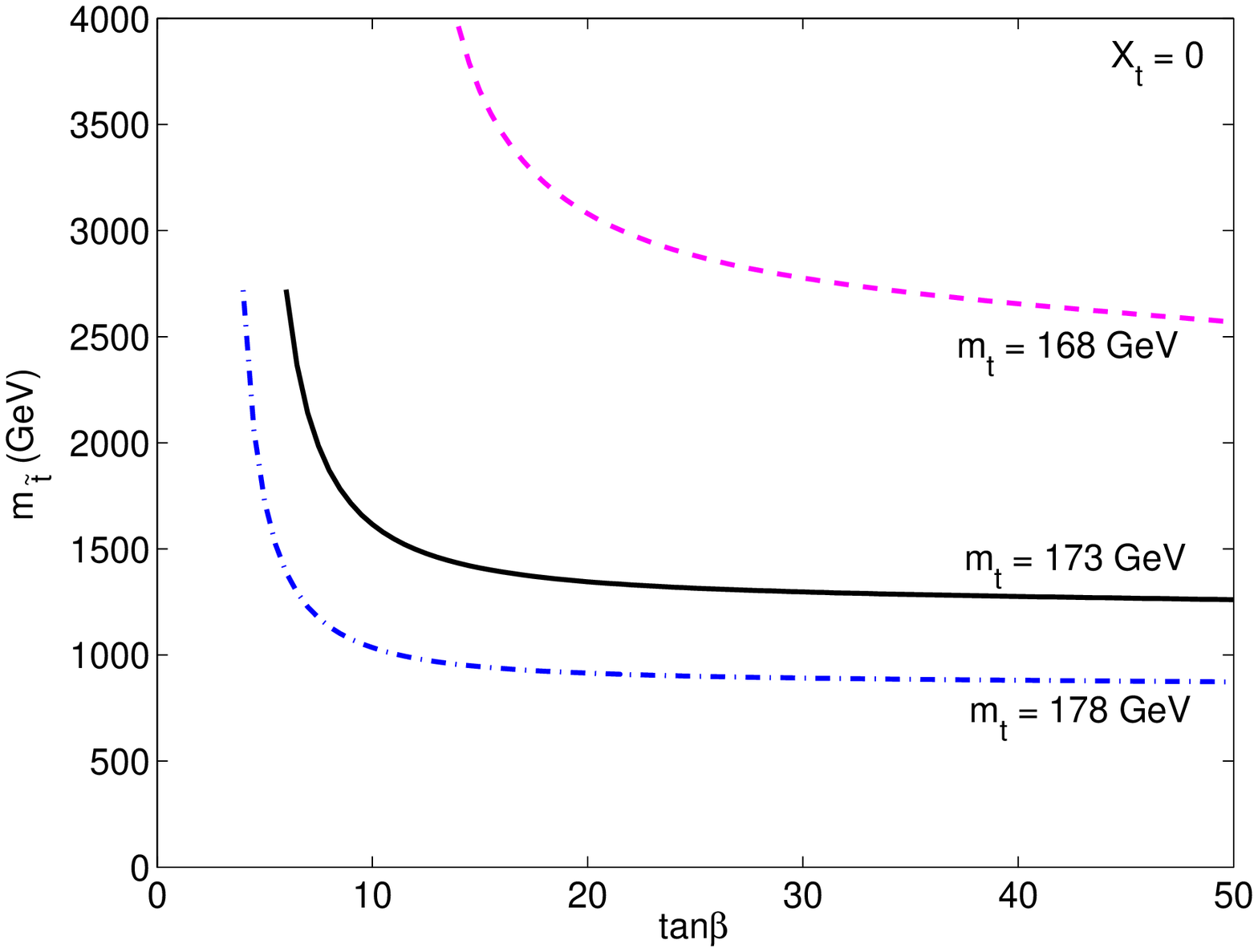}
\caption{Minimum stop soft masses, $m_{\tilde{t}} \equiv
m_{\tilde{t}_L} = m_{\tilde{t}_R} $, for $m_h \geq 114.4$ GeV as a
function of $\tan \beta$ for vanishing stop mixing ($X_t = 0$) for a
top quark mass of $m_t = 168, 173, 178$ GeV. Other parameters are as
given in Fig.~\ref{Fig:M_S_vs_TB_114_mtop173_all_mixings}.
\label{Fig:M_S_vs_TB_Xt_0_all_mtops}}
\end{center}\end{figure}

\begin{figure}\begin{center}\includegraphics[scale=0.47]{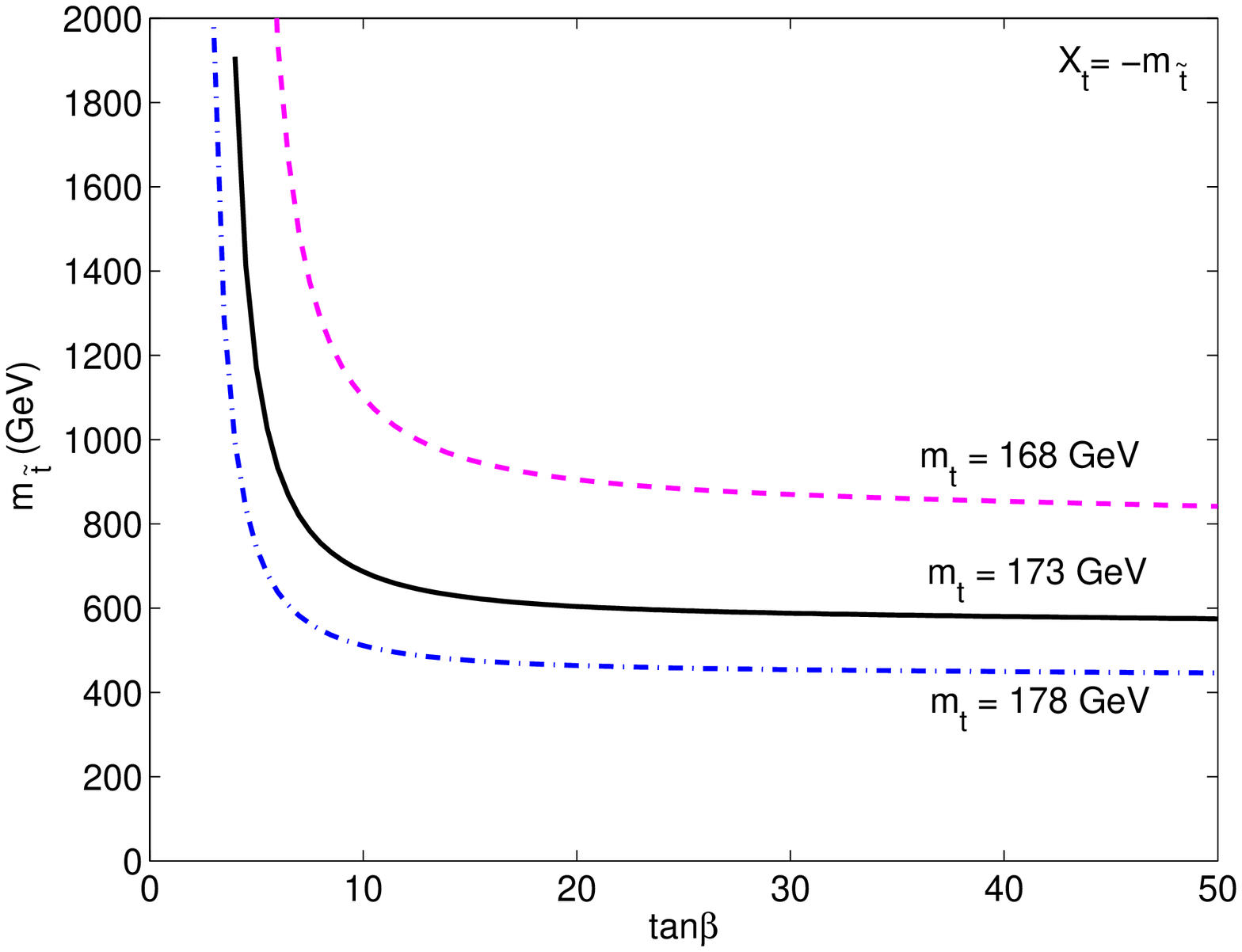}
\caption{Minimum stop soft masses, $m_{\tilde{t}} \equiv
m_{\tilde{t}_L} = m_{\tilde{t}_R} $, for $m_h \geq 114.4$ GeV as a
function of $\tan \beta$ for intermediate stop mixing ($X_t/
m_{\tilde{t}}= -1$) for a top quark mass of $m_t = 168, 173, 178$
GeV. Other parameters are as given in
Fig.~\ref{Fig:M_S_vs_TB_114_mtop173_all_mixings}.
\label{Fig:M_S_vs_TB_Xt_minus1_all_mtops}}
\end{center}\end{figure}

\begin{figure}\begin{center}\includegraphics[scale=0.47]{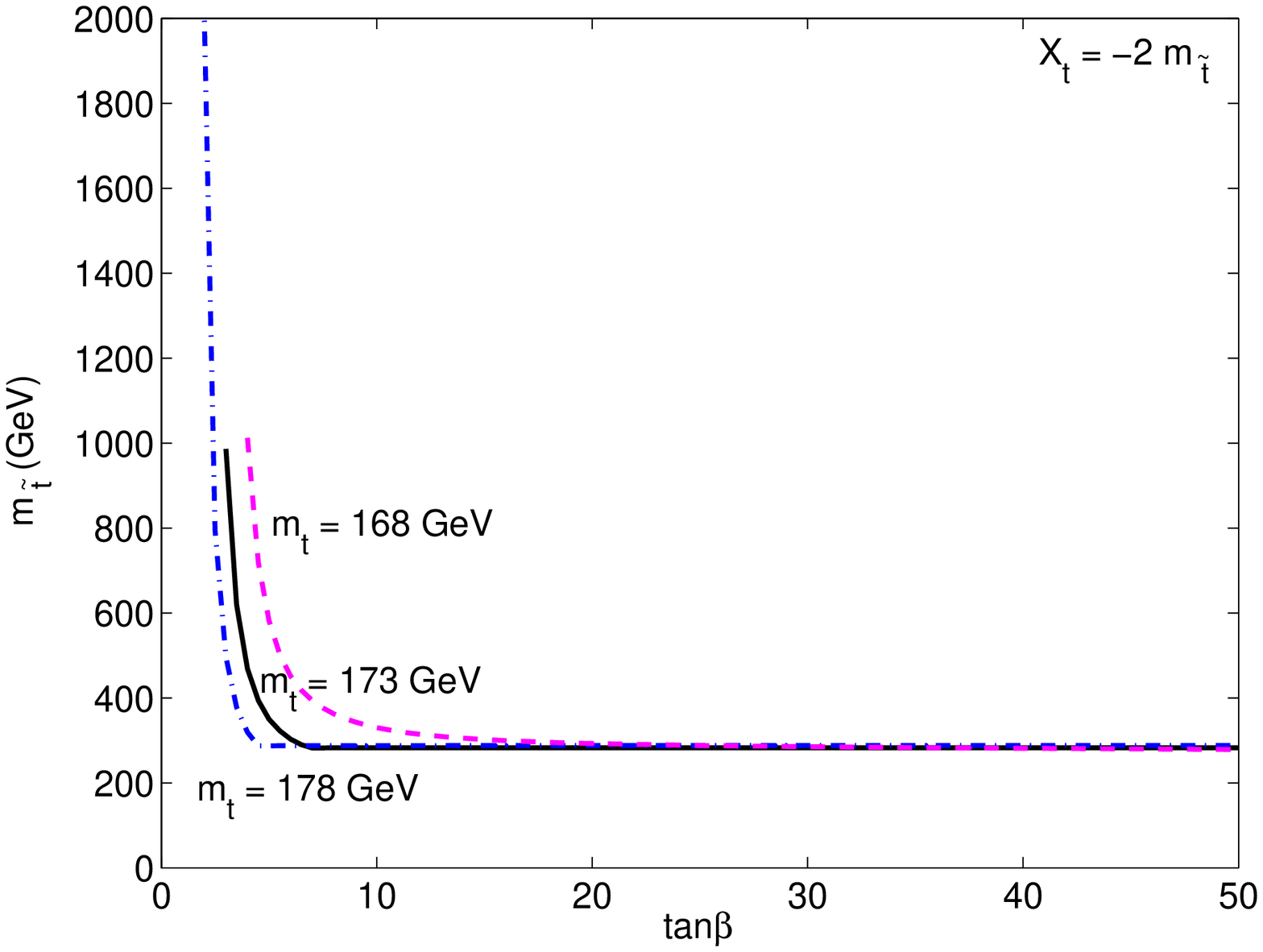}
\caption{Minimum stop soft masses, $m_{\tilde{t}} \equiv
m_{\tilde{t}_L} = m_{\tilde{t}_R} $, for $m_h \geq 114.4$ GeV as a
function of $\tan \beta$ for natural maximal stop mixing ($X_t/
m_{\tilde{t}}= -2$) for a top quark mass of $m_t = 168, 173, 178$
GeV. Other parameters are as given in
Fig.~\ref{Fig:M_S_vs_TB_114_mtop173_all_mixings}.
\label{Fig:M_S_vs_TB_Xt_minus2_all_mtops}}
\end{center}\end{figure}

The current value of the top mass from the CDF and D0 experiments at
Fermilab is $m_t = 171.4 \pm 2.1$ GeV \cite{Group:2006xn}.  The
values obtained for $\MSusyMin$ are, however, extremely sensitive to
slight variations in the value of the top mass (see also
\cite{Kitano:2006gv}). It is thus illustrative to plot $\MSusyMin$
as a function of $\tan\beta$ for various amounts of stop mixing and
for three choices of the top mass: 168 GeV, 173 GeV and 178 GeV. The
plots are shown in Fig.~\ref{Fig:M_S_vs_TB_Xt_0_all_mtops},
\ref{Fig:M_S_vs_TB_Xt_minus1_all_mtops} and
\ref{Fig:M_S_vs_TB_Xt_minus2_all_mtops} for $X_t/\MSusy$ = 0, -1,
and -2, respectively. These plots again assume that all squark,
slepton, and gaugino soft masses are equal to $\MSusy$, $\mu=200$
GeV, and all the soft trilinear couplings are equal to $A_t$.

All three figures show that $\MSusyMin$ is extremely sensitive to
small changes in $m_t$ for small $\tan\beta$.  For intermediate and
vanishing stop mixing, this sensitivity persists for large
$\tan\beta$. For example, in the no-mixing case for $\tan\beta=50$,
we find $\MSusyMin\simeq$ 870 GeV, 1260 GeV, and 2570 GeV for $m_t =
$ 178 GeV, 173 GeV, and 168 GeV, respectively. The very large value
of $\MSusyMin$ for $m_t = 168$ GeV is particularly noteworthy,
especially if the central value of the measured top mass keeps
decreasing slightly as more data from the Tevatron becomes
available.

So far we assumed that the gaugino masses are all equal to $\MSusy$.
The bino and wino masses, $M_1$ and $M_2$, as well as $\mu$
contribute to the Higgs masses at one loop, whereas the gluino mass,
$M_3$, only appears at two loops (see e.g.
\cite{Haber:1996fp,Heinemeyer:1998np,Haber:1993an} and references
therein).  Since large values of $M_1$, $M_2$ and $\mu$ can give
important \emph{negative} contributions to the Higgs masses
\cite{Carena:1999xa}, smaller values of $\MSusyMin$
are possible for smaller values of $M_1$, $M_2$ and $\mu$. For
example, setting $M_1$ = 100 GeV, $M_2$ = 200 GeV and $M_3$ = 800
GeV, we find in the no-mixing case for $\tan\beta=50$ that
$\MSusyMin\simeq$ 760 GeV, 980 GeV, and 1410 GeV for $m_t = $ 178 GeV,
173 GeV and 168 GeV, respectively. This may be compared with the values given in the
previous paragraph for the case where all the gaugino masses are
equal to $\MSusy$. Thus, setting the bino and wino masses to smaller values
decreases the size of $\MSusyMin$, especially if $m_t$ is
small. However, the large value of $\MSusyMin$ for $m_t =
168$ GeV is still noteworthy.

\begin{figure}\begin{center}\includegraphics[scale=0.47]{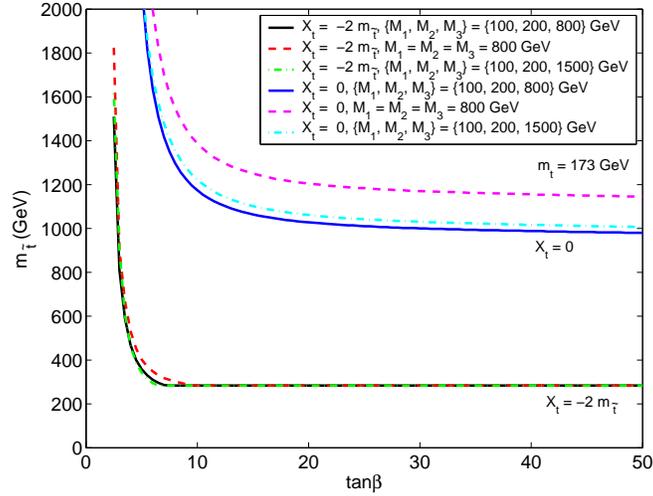}
\caption{Minimum stop soft masses, $m_{\tilde{t}} \equiv
m_{\tilde{t}_L} = m_{\tilde{t}_R} $, for $m_h \geq 114.4$ GeV as a
function of $\tan \beta$ for various values of the bino, wino, and
gluino soft masses, $M_1, M_2, M_3$. The upper three lines are for
vanishing stop mixing ($X_t=0$) and the lower three for natural
maximal stop mixing ($X_t/ m_{\tilde{t}}= -2$). Other parameters are
as given in Fig.~\ref{Fig:M_S_vs_TB_114_mtop173_all_mixings}.
\label{Fig:M_S_vs_TB_different_gaug_M1M2M3_Xt_0_and_minus2}}
\end{center}\end{figure}

We show a further example of how a different choice for $M_1$ and
$M_2$ affects $\MSusyMin$ in
Fig.~\ref{Fig:M_S_vs_TB_different_gaug_M1M2M3_Xt_0_and_minus2} for
the no-mixing and natural-maximal-mixing scenario.  For each
scenario, this figure shows a case for which $M_1$ and $M_2$ are
both large ($M_1 = M_2 = $ 800 GeV) or both small ($M_1$ = 100 GeV,
$M_2$ = 200 GeV).  In both cases, $M_3$ is fixed to be 800 GeV, $\mu
= 200$ GeV, $m_t = 173$ GeV, all squark and slepton soft masses are
equal to the stop soft masses, and all the soft trilinear couplings
are equal to $A_t$. The plots show that $\MSusyMin$ is smaller for
smaller values of $M_1$ and $M_2$.  For example, $\MSusyMin$ is
about 160 GeV smaller in the no-mixing case for $\tan\beta=50$ when
choosing the smaller set of values for $M_1$ and $M_2$, and the
difference in $\MSusyMin$ grows as $\tan\beta$ decreases. For
natural maximal mixing, no difference can be seen for most
$\tan\beta$ values, since here the condition $m_{\tilde{t}_1} \ge
100$ GeV again requires larger values of $\MSusyMin$ than the
condition $m_h \ge 114.4$ GeV. However, there is a difference in
$\MSusyMin$ for smaller $\tan\beta$, which again grows as
$\tan\beta$ decreases.

Fig.~\ref{Fig:M_S_vs_TB_different_gaug_M1M2M3_Xt_0_and_minus2} also
shows how a change in the gluino mass, $M_3$, affects $\MSusyMin$.
In general, $m_h$ tends to be maximized for $M_3$ $\simeq$ $0.8
\MSusy$ \cite{Heinemeyer:1998np}. In this figure, we compare
$\MSusyMin$ for two different gluino masses, namely $M_3 = 800$ GeV
and $M_3 = 1500$ GeV.  The figure shows that the effect is not very
large for this choice of parameters.  However, the gluino mass can
significantly affect the Higgs masses, and therefore $\MSusyMin$,
for large $\tan\beta$ and large and \emph{negative} $\mu$.

The variation of $m_h$ as a function of $\mu$ does not generally
exceed about 3 GeV \cite{Heinemeyer:1998np}. However, it can become
very large if one includes the all-order resummation of the
$\tan\beta$ enhanced terms of order $\mathcal{O}(\alpha_b (\alpha_s
\tan\beta)^n)$, where $\alpha_b = \lambda_b^2/4\pi$ and $\lambda_b$
is the bottom Yukawa coupling
\cite{Banks:1987iu,Hall:1993gn,Hempfling:1993kv,Carena:1994bv,Carena:1999py,Eberl:1999he,
Heinemeyer:2004xw}. This resummation is included in
$\mathtt{FeynHiggs}$. The origin of the enhancement is a change in
the bottom Yukawa coupling due to a loop containing, for example, a
gluino and a sbottom squark. The leading corrections to the bottom
Yukawa coupling can be incorporated into the one-loop result for the
Higgs masses by the use of an effective bottom mass,
$m_b^{\textrm{\tiny{eff}}}$. Large $|\mu|\tan\beta$ can
substantially change the effective bottom mass
$m_b^{\textrm{\tiny{eff}}}$ from its $\overline{\textrm{MS}}$ value.
\emph{Positive} $\mu$ can substantially \emph{decrease}
$m_b^{\textrm{\tiny{eff}}}$, making the sbottom/bottom sector
corrections to $m_h$ negligible. \emph{Negative} $\mu$ on the other
hand can substantially \emph{increase} $m_b^{\textrm{\tiny{eff}}}$,
making the sbottom/bottom sector corrections to $m_h$ important. The
bottom/sbottom corrections to $m_h$ are negative in the latter case.
Larger stop masses are then required for large and negative $\mu$ as
$\tan\beta$ increases to enhance the positive radiative corrections
from the stop/top.

\begin{figure}\begin{center}\includegraphics[scale=0.47]{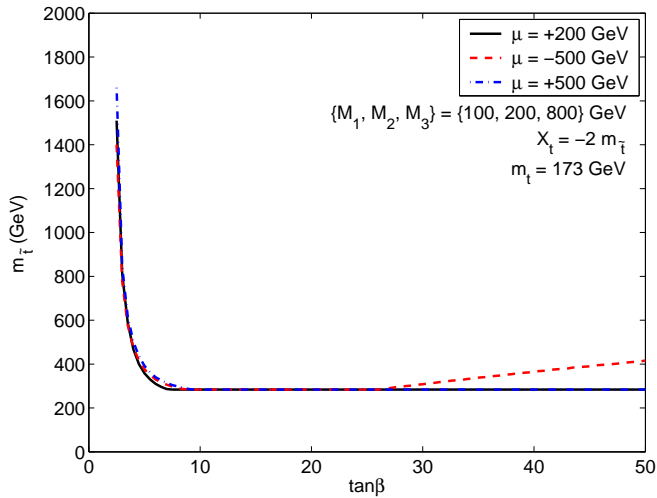}
\caption{Minimum stop soft masses, $m_{\tilde{t}} \equiv
m_{\tilde{t}_L} = m_{\tilde{t}_R} $, for $m_h \geq 114.4$ GeV as a
function of $\tan \beta$ for natural maximal stop mixing ($X_t/
m_{\tilde{t}}= -2$) with $\mu=-500,200,500$ GeV and bino, wino, and
gluino soft masses of $M_1 = 100$ GeV, $M_2=200$ GeV, $M_3=800$ GeV.
Other parameters are as given in
Fig.~\ref{Fig:M_S_vs_TB_114_mtop173_all_mixings}.
\label{Fig:M_S_vs_TB_different_mu_Xt_minus2}}
\end{center}\end{figure}

This effect can be seen in
Fig.~\ref{Fig:M_S_vs_TB_different_mu_Xt_minus2} where we compare
$\mu = +200$ GeV and $\mu = \pm 500$ GeV for natural maximal stop
mixing. This figure again assumes that all squark, slepton, and
gaugino soft masses are equal to the stop soft masses, $m_t = 173$
GeV, and all the soft trilinear couplings are equal to $A_t$. For
large $\tan\beta$, slightly larger $\MSusyMin$ are required for $\mu
= -500$ GeV than when $\mu$ is positive (the effect would be
stronger for even larger negative $\mu$). Note that for small values
of $\tan\beta$ there is a region for which $\MSusy$ is larger for
both $\mu = -500$ GeV and $\mu = +500$ GeV than for $\mu = 200$ GeV.
As we discussed above, this is because larger chargino and
neutralino masses decrease the size of $m_h$.

Since the gluino mass also enters the equation that determines
$m_b^{\textrm{\tiny{eff}}}$, it can have a significant impact on
$m_h$ for large $\tan\beta$ and large negative values of $\mu$ as
demonstrated in \cite{Heinemeyer:2004xw}. Thus, some non-negligible
dependence of $\MSusyMin$ on the gluino mass is expected for
negative and large $\mu$.

\subsection{Lower bounds on the stop masses as function of the Higgs mass}\label{Sec:various mh}

In this section, we present lower bounds on the stop soft masses,
$m_{\tilde{t}} \equiv m_{\tilde{t}_L} = m_{\tilde{t}_R}$, as a
function of the Higgs mass, $m_h$.  We assume the decoupling limit
($m_A = 1000$ GeV), and we set all squark and slepton soft masses
equal to the stop soft masses, $\mu=200$ GeV, $m_t = 173$ GeV, $M_1$
= 100 GeV, $M_2$ = 200 GeV, $M_3$ = 800 GeV, $\tan\beta = 30$, and
all the soft trilinear couplings equal to $A_t$.  We allow $m_h$ to
range from 100 GeV upwards.   This means that the values obtained
for $\MSusy$ in the range $m_h \in [100,114.4]$ GeV will be lower
than those consistent with the LEP results, since we set no
additional constraints on $\cossqbma$ and $m_H$. However, the main
point here is to show the dependence of $\MSusy$ on $m_h$ without
any other constraints. The lower bounds on $\MSusy$ are required to
give physical stop masses not less than 100~GeV.

\begin{figure}\begin{center}\includegraphics[scale=0.50]{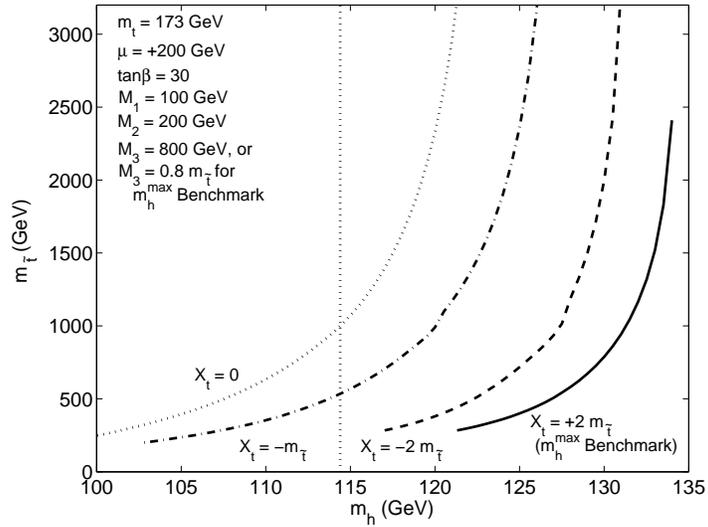}
\caption{Minimum stop soft masses, $m_{\tilde{t}} \equiv
m_{\tilde{t}_L} = m_{\tilde{t}_R} $, as a function of $m_h$.  All
squark and slepton soft masses are equal to the stop soft masses,
$\mu=200$ GeV, $m_t = 173$ GeV, $\{M_1,M_2\}$ = $\{100, 200\}$ GeV,
$\tan\beta = 30$, and all the soft trilinear couplings are equal to
$A_t=X_t+\mu\cot\beta$.  In the figure, the curved lines from left
to right are as follows: the dotted line is for no mixing
($X_t/\MSusy = 0$), the dash-dot line for intermediate mixing
($X_t/\MSusy = -1$), the dashed line for natural maximal mixing
($X_t/\MSusy = -2$), and the solid line for the
$m_h^{\textrm{\tiny{max}}}$ benchmark scenario ($X_t/\MSusy = +2$)
\cite{Carena:1999xa}. The gluino mass is set to be $M_3$ = 800 GeV
in all cases except in the $m_h^{\textrm{\tiny{max}}}$ benchmark
scenario, where $M_3 = 0.8\MSusy$. The vertical dotted line is at
$m_h = 114.4$ GeV, which is the lower bound set by LEP on $m_h$ in
the decoupling limit. \label{Fig:M_S_vs_m_h_various_mixings}}
\end{center}\end{figure}

We show the results for different amounts of stop mixing in
Fig.~\ref{Fig:M_S_vs_m_h_various_mixings}. This figure shows how an
increase in $m_h$ requires an exponential increase in $\MSusy$. In
addition to the no-mixing, intermediate-mixing and
natural-maximal-mixing cases, we also include the
$m_h^{\textrm{\tiny{max}}}$ benchmark scenario ($X_t = +2\MSusy$)
(but with $\mu = +200$ GeV, not $\mu = -200$ GeV)
\cite{Carena:1999xa}. This benchmark scenario is designed to
maximize the Higgs mass for a given set of parameters. Moreover, we
choose $M_3 = 800$ GeV for all cases, with the exception of the
latter one. In the latter benchmark scenario, we choose the
benchmark value $M_3 = 0.8 \MSusy$ instead, which gives slightly
higher values for $m_h$ \cite{Carena:1999xa}.

It is clear from the figure that there is some value of $m_h$ at
which a further small increase in $m_h$ would require an extremely
large increase in the stop masses.  It is instructive to obtain the
values of $m_h$ from the figure if, for example, $\MSusy$ = 3000
GeV.  We find for no stop mixing, $m_h \simeq 121$ GeV, for
intermediate stop mixing, $m_h \simeq 126$ GeV, for natural maximal
stop mixing, $m_h \simeq 131$ GeV, and for the
$m_h^{\textrm{\tiny{max}}}$ benchmark scenario, $m_h \simeq 134$ GeV
(see also \cite{Carena:2002es}, for example, and references
therein).

Since $A_t$ and $M_3$ most naturally have the opposite sign due to
renormalization group running and the presence of a strongly
attractive quasi-fixed point (see Appendix \ref{Sec:At fixed
point}), a negative value of $A_t$ is more natural. For negative
$A_t$, the upper bound of $m_h$ in the MSSM is around 131 GeV.

\subsection{Lower bounds on the stop masses for $m_h \simeq 93$ GeV}\label{Sec:m_h_simeq_93_GeV}

In this section, we present results for the minimum stop soft
masses, $m_{\tilde{t}} \equiv m_{\tilde{t}_L} = m_{\tilde{t}_R}$, as
a function of $\tan\beta$, for various choices of the other MSSM
parameters, and consistent with the following set of constraints on
the Higgs sector obtained by LEP: $m_h\simeq 93$ GeV, $\cossqbma \ge
0.8$ and $m_H \ge 114.4$ GeV (see equation (\ref{Eqn:lower Region
B})).

The mass $m_A$ is allowed to be a free parameter, since $\MSusy$
needs to be minimized without enforcing the decoupling limit. We
vary $m_A$ between 93.5 GeV and 1000 GeV from the bottom up for a
given choice of $\MSusy$ and other MSSM parameters, until the
conditions 93 GeV $\le$ $m_h$ $\le$ 95 GeV, $\cossqbma \ge 0.8$, and
$m_H \ge 114.4$ GeV are satisfied. (The lower bound of 93.5 GeV for
$m_A$ is the approximate lower bound obtained within the same
benchmark scenarios as the bound on $m_h$; it turns out that the
actual values obtained for $m_A$ are slightly larger). If these
conditions cannot all be satisfied, we keep increasing $\MSusy$
until they are satisfied.  Note that we require the lower bounds on
$\MSusy$ to give physical stop masses of at least 100 GeV. We again
denote the lower bounds on $\MSusy$ consistent with the LEP Higgs
bounds by $\MSusyMin$.

The Higgs masses, $m_h$ and $m_H$, are calculated with
$\mathtt{FeynHiggs}$, and $\cossqbma$ is calculated using the
$\mathtt{FeynHiggs}$ output of the radiatively corrected CP-even
Higgs mixing angle $\alpha$.

\begin{figure}[!bht]\begin{center}\includegraphics[scale=0.47]{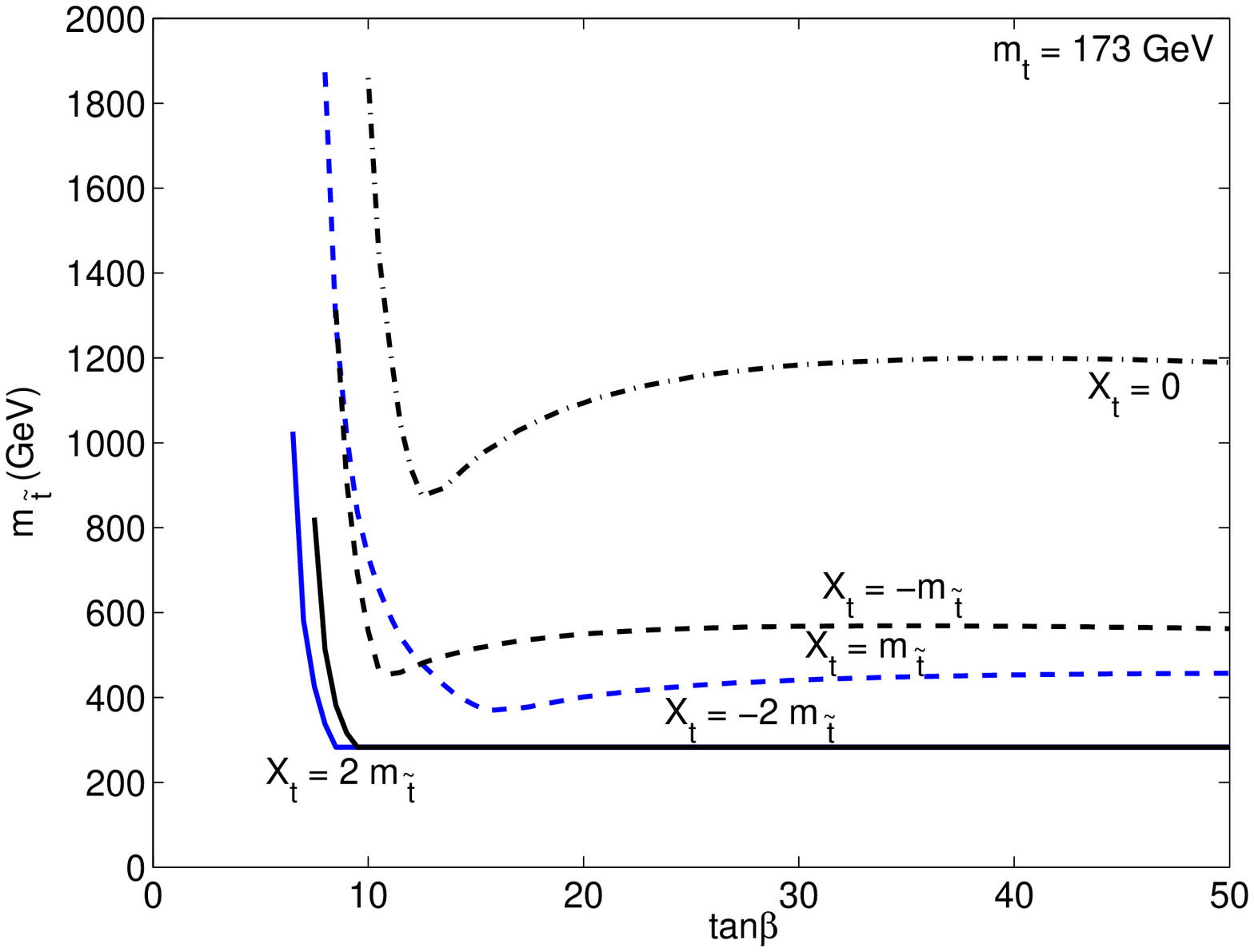}
\caption{Minimum stop soft masses, $m_{\tilde{t}} \equiv
m_{\tilde{t}_L} = m_{\tilde{t}_R} $, for $m_h\simeq 93$ GeV, $m_H
\ge 114.4$ GeV, and $\cossqbma \ge 0.8$, as a function of
$\tan\beta$ for stop mixing $X_t/ m_{\tilde{t}} = 0,\pm 1, \pm 2$.
All squark, slepton, and gaugino soft mass parameters are equal to
the stop soft masses, $\mu=200$ GeV, $m_t = 173$ GeV, and all soft
trilinear couplings are equal to $A_t=X_t+\mu\cot\beta$.  This
figure may be compared with
Fig.~\ref{Fig:M_S_vs_TB_114_mtop173_all_mixings}.
\label{Fig:M_S_vs_TB_tongue_mtop173_all_mixings}}
\end{center}\end{figure}

In Fig.~\ref{Fig:M_S_vs_TB_tongue_mtop173_all_mixings}, we show
$\MSusyMin$ as a function of $\tan\beta$ for stop mixing $X_t/
m_{\tilde{t}}$ $=$ $0$, $\pm 1$, and $\pm 2$.  All squark, slepton,
and gaugino soft masses are equal to the stop soft masses, $\mu=200$
GeV, $m_t = 173$ GeV, and all the soft trilinear couplings are equal
to the stop soft trilinear coupling, $A_t$. This figure may be
compared with Fig.~\ref{Fig:M_S_vs_TB_114_mtop173_all_mixings} in
which we require $m_h \ge 114.4$ GeV in the Higgs decoupling limit.

\begin{figure}\begin{center}\includegraphics[scale=0.47]{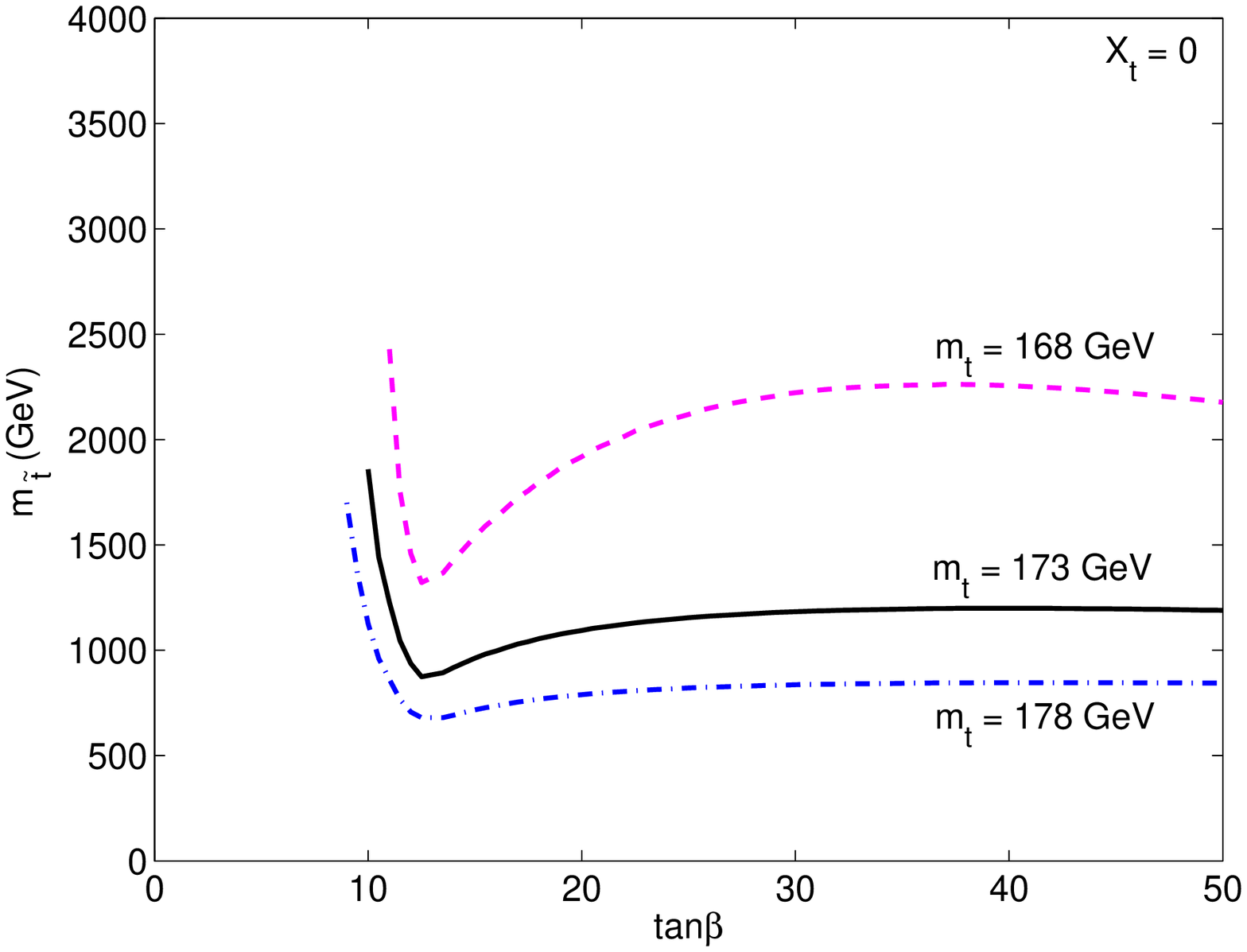}
\caption{Minimum stop soft masses, $m_{\tilde{t}} \equiv
m_{\tilde{t}_L} = m_{\tilde{t}_R} $, for $m_h\simeq 93$ GeV, $m_H
\ge 114.4$ GeV, and $\cossqbma \ge 0.8$, as a function of
$\tan\beta$ for vanishing stop mixing ($X_t/ m_{\tilde{t}} = 0$) for
a top quark mass of $m_t = 168, 173, 178$ GeV. Other parameters are
given as in Fig.~\ref{Fig:M_S_vs_TB_tongue_mtop173_all_mixings}.
This figure may be compared with
Fig.~\ref{Fig:M_S_vs_TB_Xt_0_all_mtops}.
\label{Fig:M_S_vs_TB_tongue_Xt_0_all_mtops}}
\end{center}\end{figure}

\begin{figure}\begin{center}\includegraphics[scale=0.47]{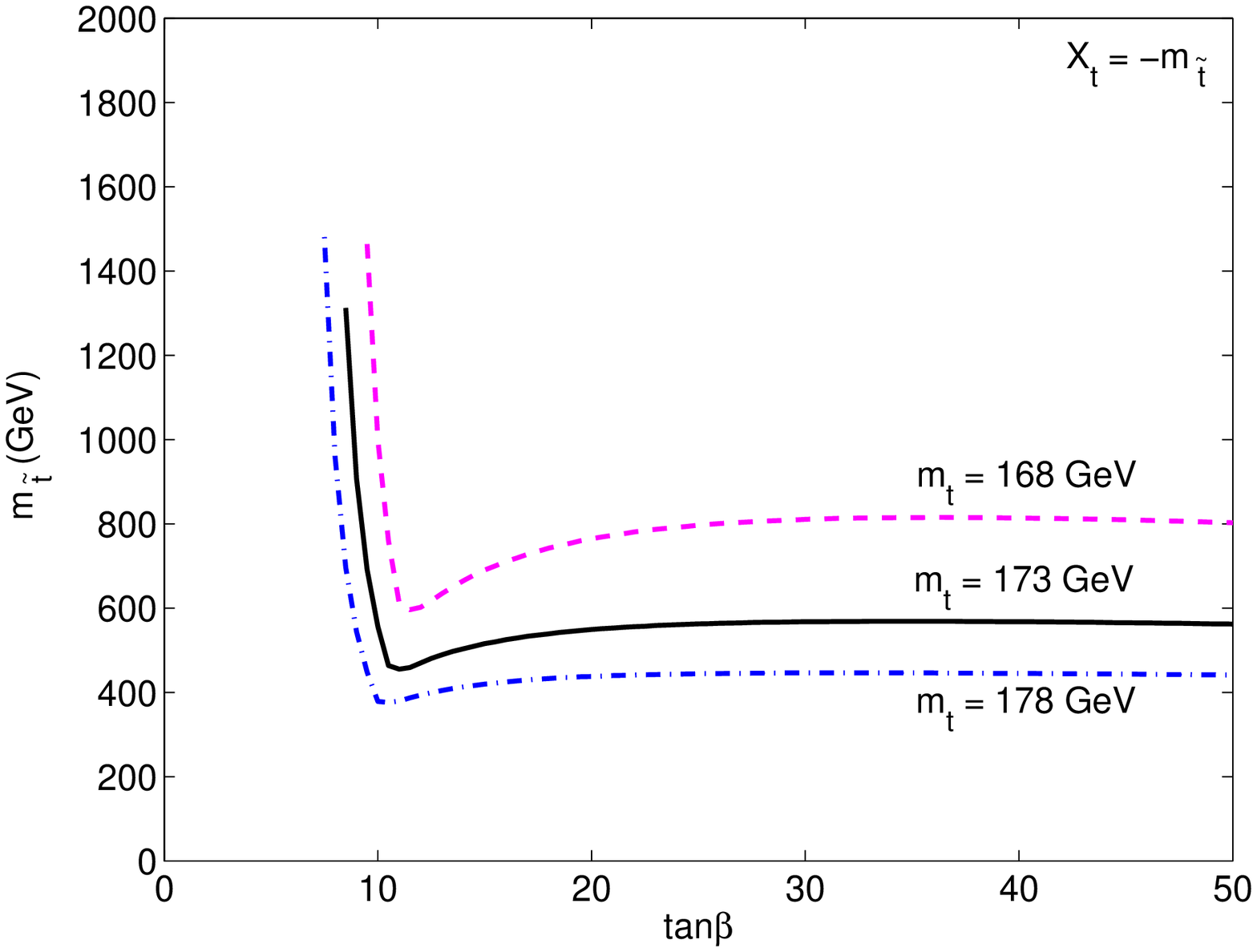}
\caption{Minimum stop soft masses, $m_{\tilde{t}} \equiv
m_{\tilde{t}_L} = m_{\tilde{t}_R} $, for $m_h\simeq 93$ GeV, $m_H
\ge 114.4$ GeV, and $\cossqbma \ge 0.8$, as a function of
$\tan\beta$ for intermediate stop mixing ($X_t/ m_{\tilde{t}} = -1$)
for a top quark mass of $m_t = 168, 173, 178$ GeV. Other parameters
are given as in Fig.~\ref{Fig:M_S_vs_TB_tongue_mtop173_all_mixings}.
This figure may be compared with
Fig.~\ref{Fig:M_S_vs_TB_Xt_minus1_all_mtops}.
\label{Fig:M_S_vs_TB_tongue_Xt_minus1_all_mtops}}
\end{center}\end{figure}

\begin{figure}\begin{center}\includegraphics[scale=0.4]{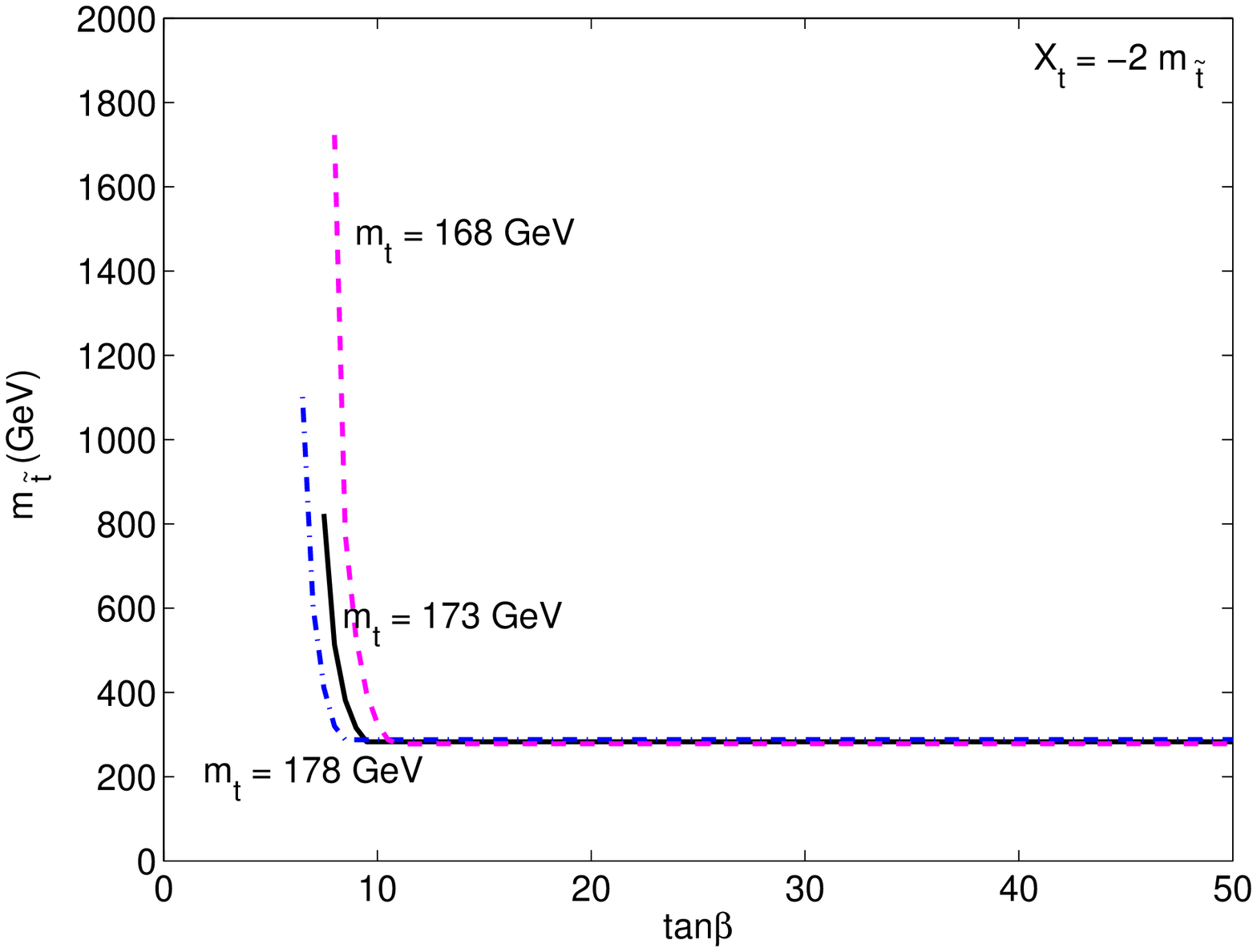}
\caption{Minimum stop soft masses, $m_{\tilde{t}} \equiv
m_{\tilde{t}_L} = m_{\tilde{t}_R} $, for $m_h\simeq 93$ GeV, $m_H
\ge 114.4$ GeV, and $\cossqbma \ge 0.8$, as a function of
$\tan\beta$ for natural maximal stop mixing ($X_t/ m_{\tilde{t}} =
-2$) for a top quark mass of $m_t = 168, 173, 178$ GeV. Other
parameters are given as in
Fig.~\ref{Fig:M_S_vs_TB_tongue_mtop173_all_mixings}. This figure may
be compared with Fig.~\ref{Fig:M_S_vs_TB_Xt_minus2_all_mtops}.
\label{Fig:M_S_vs_TB_tongue_Xt_minus2_all_mtops}}
\end{center}\end{figure}

Next, we show $\MSusyMin$ as a function of $\tan\beta$ for different
values of the top mass (168 GeV, 173 GeV and 178 GeV) and for
different amounts of mixing.
Fig.~\ref{Fig:M_S_vs_TB_tongue_Xt_0_all_mtops} is for no mixing,
Fig.~\ref{Fig:M_S_vs_TB_tongue_Xt_minus1_all_mtops} is for
intermediate mixing, and
Fig.~\ref{Fig:M_S_vs_TB_tongue_Xt_minus2_all_mtops} is for natural
maximal mixing.  All squark, slepton, and gaugino soft masses are
again equal to the stop soft masses, $\mu=200$ GeV, and all the soft
tri-linear couplings are equal to $A_t$.  These figures may be
compared with the figures in which we require $m_h \ge 114.4$ GeV in
the decoupling limit, namely Figs.
\ref{Fig:M_S_vs_TB_Xt_0_all_mtops},
\ref{Fig:M_S_vs_TB_Xt_minus1_all_mtops} and
\ref{Fig:M_S_vs_TB_Xt_minus2_all_mtops}, respectively.

We first compare $\MSusyMin$ in the two scenarios $m_h \simeq 93$
GeV and $m_h \ge 114.4$ GeV for large $\tan\beta$.  Here, the
figures show that $\MSusyMin$ is the same in the case of maximal or
natural maximal mixing. For intermediate and vanishing stop mixing,
$\MSusyMin$ is only slightly smaller for $m_h \simeq 93$ GeV than
for $m_h \ge 114.4$ GeV. Assuming $m_t$ = 173 GeV and $\tan\beta =
50$, the difference is only about 15 GeV for $X_t$ = $-\MSusy$ and
70 GeV for $X_t$ = 0. We expected the values for $\MSusyMin$ to be
so similar from the discussion in Section \ref{Sec:Implication for
MSSM stop sector}.

For moderate $\tan\beta$, $\MSusyMin$ can be substantially smaller for
$m_h \simeq 93$ GeV than for $m_h \ge 114.4$ GeV. This is true in
particular for the no-mixing and intermediate-mixing cases, with
the difference being more pronounced for smaller values of $m_t$.
For example, the maximum difference between $\MSusyMin$ in the two
scenarios is about 600 GeV for $\tan\beta = 12.5$ if there is no
mixing and $m_t=173$ GeV.

As $\tan\beta$ decreases further, however, $\MSusyMin$ for $m_h \simeq
93$ GeV rises very steeply, and becomes larger than for $m_h \ge
114.4$ GeV.

\begin{figure}\begin{center}\includegraphics[scale=0.4]{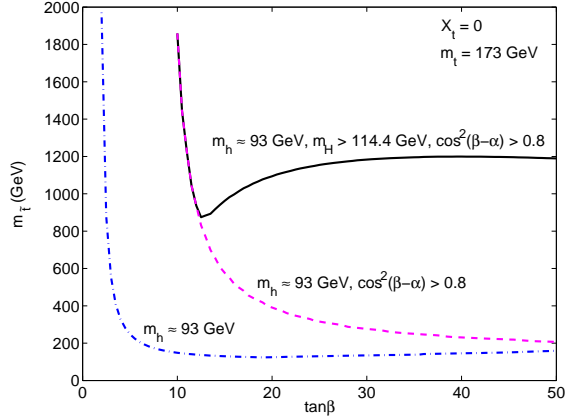}
\caption{Minimum stop soft masses, $m_{\tilde{t}} \equiv
m_{\tilde{t}_L} = m_{\tilde{t}_R} $, as a function of $\tan\beta$,
for no stop mixing ($X_t/ m_{\tilde{t}} = 0$).  The solid line shows
the minimum stop soft masses for $m_h \simeq 93$ GeV, $m_H \ge
114.4$ GeV, and $\cos^2(\beta-\alpha) \ge 0.8$, consistent with the
LEP Higgs bounds. The dashed and dash-dot lines are not consistent
with the LEP bounds and used for comparison. The dashed line shows
the minimum soft masses for $m_h \simeq 93$ GeV and
$\cos^2(\beta-\alpha) \ge 0.8$ and without a constraint on $m_H$.
The dash-dot line shows the minimum soft masses for $m_h \simeq 93$
GeV and without constraints on $m_H$ and $\cos^2(\beta-\alpha)$.
Other parameters are given as in
Fig.~\ref{Fig:M_S_vs_TB_tongue_mtop173_all_mixings}.
\label{Fig:M_S_vs_TB_tongue_no114_93_Xt_0}}
\end{center}\end{figure}

\begin{figure}\begin{center}\includegraphics[scale=0.47]{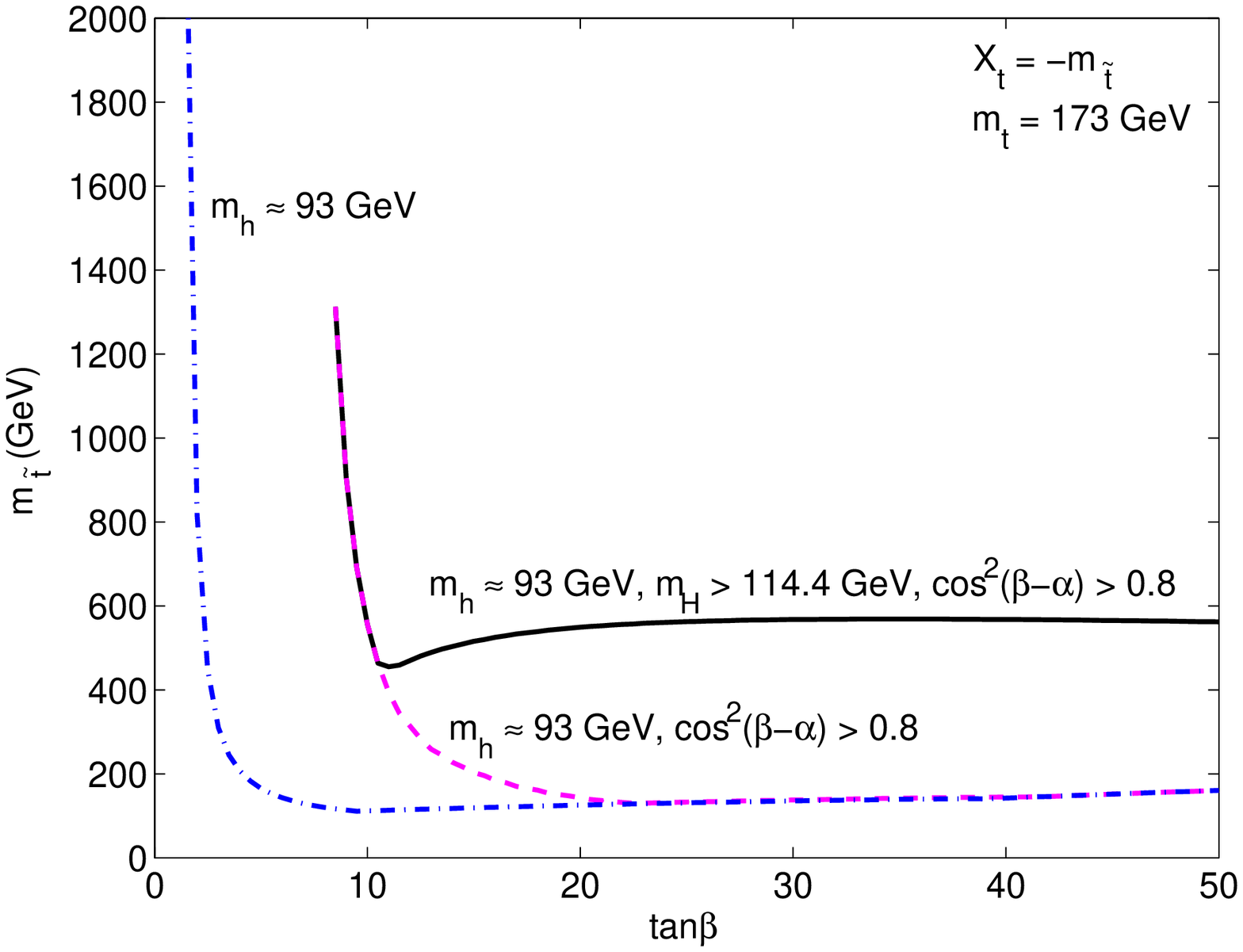}
\caption{Minimum stop soft masses, $m_{\tilde{t}} \equiv
m_{\tilde{t}_L} = m_{\tilde{t}_R} $, as a function of $\tan\beta$,
for intermediate stop mixing ($X_t/ m_{\tilde{t}} = -1$).  The other
parameters and the different lines are as for
Fig.~\ref{Fig:M_S_vs_TB_tongue_no114_93_Xt_0}.
\label{Fig:M_S_vs_TB_tongue_no114_93_Xt_minus1}}
\end{center}\end{figure}

\begin{figure}\begin{center}\includegraphics[scale=0.47]{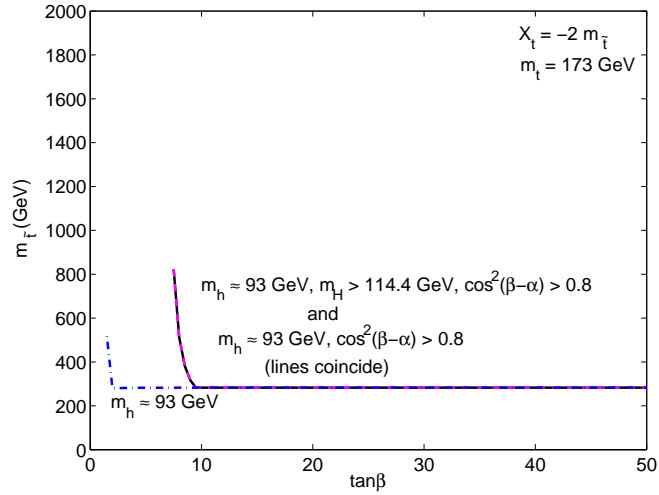}
\caption{Minimum stop soft masses, $m_{\tilde{t}} \equiv
m_{\tilde{t}_L} = m_{\tilde{t}_R} $, as a function of $\tan\beta$,
for natural maximal stop mixing ($X_t/ m_{\tilde{t}} = -2$). The
other parameters and the different lines are as for
Fig.~\ref{Fig:M_S_vs_TB_tongue_no114_93_Xt_0}.
\label{Fig:M_S_vs_TB_tongue_no114_93_Xt_minus2}}
\end{center}\end{figure}

Understanding this behavior of $\MSusyMin$ as a function of
$\tan\beta$ requires an understanding of the importance of the
constraints $\cossqbma\ge 0.8$ and $m_H\ge 114.4$ GeV.  To this end,
we compare $\MSusyMin$ versus $\tan\beta$ for the case that the
constraint on $m_H$ is ignored, for the case that both constraints
are ignored, and for the case consistent with the LEP bounds that
includes both constraints. We again make the comparison for various
amounts of mixing in the stop sector.
Fig.~\ref{Fig:M_S_vs_TB_tongue_no114_93_Xt_0} shows the results for
no mixing, Fig.~\ref{Fig:M_S_vs_TB_tongue_no114_93_Xt_minus1} for
intermediate mixing, and
Fig.~\ref{Fig:M_S_vs_TB_tongue_no114_93_Xt_minus2} for natural
maximal mixing.  Each of these figures has three lines. The solid
line shows the results which are consistent with the LEP bounds,
i.e. it includes the two constraints $\cos^2(\beta-\alpha) \ge 0.8$
and $m_H \ge 114.4$ GeV, in addition to requiring $m_h \simeq 93$
GeV. The dashed line, on the other hand, does not include the
constraint on $m_H$, but does require $\cos^2(\beta-\alpha)\ge 0.8$
and $m_h \simeq 93$ GeV.  The dash-dot line only requires $m_h
\simeq 93$ GeV, and ignores the constraints on
$\cos^2(\beta-\alpha)$ and $m_H$.

As expected, both constraints from LEP in general increase
$\MSusyMin$.  The constraint $\cossqbma \ge 0.8$ is more important
as $\tan\beta$ becomes smaller, but less important as $\tan\beta$
gets larger. The constraint $m_H \ge 114.4$ GeV, however, is more
important for larger $\tan\beta$ (if stop mixing is not too large),
but is less important as $\tan\beta$ becomes smaller. We now explain
these observations.

If the only condition is $m_h \simeq 93$
GeV, the theory tends to be in the Higgs decoupling limit where $\cossqbma$
$\rightarrow$ 0.  The reason for this is that for a given set of parameters, including a
given value of $\MSusy$, $m_h$ is maximized in the
decoupling limit. (This is also the reason why ignoring both
constraints is in general equivalent to ignoring only the
constraint $\cos^2(\beta-\alpha)$ $\ge$ 0.8 but keeping $m_H \ge
114.4$ GeV as a constraint.)
The constraint $\cossqbma \ge 0.8$, however, forces all the MSSM
Higgs masses to be quite small.  In particular, $m_A$ is forced to be
relatively small and degenerate with $m_h$, so that larger $\MSusy$
are required to obtain the same value for
$m_h$.\footnote{The results for $m_A$ for the case consistent with
the LEP results (which includes both constraints) are $m_A \in$
[96.1 GeV, 99.5 GeV] for natural maximal mixing, $m_A \in$ [94.3
GeV, 97.7 GeV] for intermediate mixing, and $m_A \in$ [95.1 GeV,
97.7 GeV] for no mixing.  When the constraint on $m_H$ is ignored,
$m_A$ lies roughly in the same range.  Note that from the
pair-production channel these values of $m_h + m_A$ give upper
bounds on $\cos^2(\beta-\alpha)$ consistent with
$\cos^2(\beta-\alpha)$ $\ge$ 0.8, depending on what one assumes for
the Higgs decay branching ratios, see \cite{ALEPH:2006cr}.}
Moreover, the maximum value reached by $\cossqbma$ decreases as
$\tan\beta$ decreases. Larger radiative corrections, in particular
larger values of $\MSusy$ or more stop mixing, can increase the
maximum value of $\cossqbma$.  However, if $\tan\beta$ decreases too far,
exponentially larger values of $\MSusy$ are required to allow
$\cossqbma$ to be greater than 0.8.

For a given set of parameters, $m_h$ in general decreases as
$\tan\beta$ decreases.  This is not the case for $m_H$, which in
general decreases as $\tan\beta$ increases.  This explains why the
constraint on $m_H$ is more important for larger values of
$\tan\beta$.  In the decoupling limit, $m_H$ is approximately
degenerate with $m_A$, and larger values of $\MSusy$ do not affect
$m_H$ much. In the non-decoupling limit, however, larger values of
$\MSusy$ can increase $m_H$. In fact, if we define
$m_h^{\textrm{\tiny{max}}}$ to be equal to $m_h$ in the decoupling
limit, then $m_H$ $\simeq$ $m_h^{\textrm{\tiny{max}}}$ for large $\tan\beta$ and
$\cossqbma \simeq 1$. This may be explained with the formula

\beq m_h^2 \sinsqbma + m_H^2 \cossqbma =
(m_h^{\textrm{\tiny{max}}})^2, \eeq

\noindent valid for large $\tan\beta$
\cite{Carena:2002es,Moreno:1996zm,Carena:1999bh}, and explains why
larger $\MSusy$ increases the value of $m_H$ in, or near, the
non-decoupling region (see also Section \ref{Sec:Implication for
MSSM stop sector}).

\vspace{2cm}

\clearpage

\clearpage

\clearpage

\section{Implications of new physics constraints for the
lower bounds on the stop masses}\label{Sec:Constraints from new
physics}

In Section \ref{Sec:Lower bounds on MSusy}, we presented lower
bounds on the stop soft masses that are consistent with the LEP
Higgs bounds (we again denote these bounds by $\MSusyMin$). In this
section, we consider additional constraints from the electroweak
$S$- and $T$-parameter and the decays $B \to X_s \gamma$ and $B_s
\to \mu^+ \mu^-$, which also constrain the Higgs and/or stop sector.
Some of these constraints may provide more stringent lower bounds on
the stop masses than those provided by the constraints from LEP on
the Higgs sector, or they might indirectly constrain the Higgs
sector more tightly than the LEP results.

\subsection{Constraints from Electroweak Precision Measurements: T- and
S-parameters}

The oblique parameters $T$ and $S$ parameterize new physics
contributions to electroweak vacuum-polarization diagrams.  They
give a good parametrization if these diagrams are the dominant
corrections to electroweak precision observables \cite{Peskin:1991sw}.
Strong constraints on these parameters already exist
\cite{PDBook,Erler:2006vt}.

The MSSM includes new $SU(2)_L$ doublets that contribute to the $T$-
and $S$-parameter (which are defined to be zero from SM
contributions alone). The $T$-parameter is a measure of how strongly
the vector part of $SU(2)_L$ is broken, and is non-zero, for
example, for heavy, non-degenerate multiplets of fermions or
scalars. The $S$-parameter is a measure of how strongly the axial
part of $SU(2)_L$ is broken, and is non-zero, for example, for
heavy, degenerate multiplets of chiral fermions \cite{PDBook}.

The main contribution in the MSSM to the $T$-parameter in general
comes from the stop/sbottom doublet \cite{Heinemeyer:2004gx}.  In
particular, large mixing in the stop and/or sbottom sectors can lead
to large differences amongst the two stop and two sbottom masses,
which gives a large contribution to the $T$-parameter.  Moreover,
for a given set of parameters and fixed $X_t/\MSusy$, decreasing
$\MSusy$ tends to increase the value of the $T$-parameter. For these
reasons the $T$-parameter could provide more stringent lower bounds
on $\MSusy$ than those coming from the LEP Higgs bounds when the
mixing in the stop sector is large, since then the stop and sbottom
masses are split by large amounts and the LEP Higgs constraints
allow for small $\MSusy$.

We estimate the $T$-parameter with version 2.2.7 of
$\mathtt{FeynHiggs}$. This program calculates $\delta\rho$ which
measures the deviation of the electroweak $\rho$-parameter from
unity.  The $T$-parameter and $\delta\rho$ are related by
$\delta\rho = \alpha T$, where $\alpha$ is the QED coupling.  All
the results presented in this paper are consistent with the
2$\sigma$ constraint on the upper bound of $\delta\rho$, namely
$\delta\rho \le 0.0026$ \cite{PDBook}. The $T$- and $S$-parameters
are correlated, so that this bound corresponds to the 2$\sigma$
bound on $T$ for $S=0$.

We find that the 2$\sigma$ constraint on $\delta\rho$ does not provide
an additional constraint on the stop masses in essentially all the
analyses presented in this paper.

For (natural) maximal stop mixing with $\MSusy=\MSusyMin$, the value
of $\delta\rho$ is not consistent with its 1$\sigma$ bound, although
it is consistent with its 2$\sigma$ bound (for intermediate and less
mixing, it is consistent also with the 1$\sigma$ bound). For
example, $\MSusyMin$ = 283 GeV for large $\tan\beta$ and natural
maximal stop mixing ($X_t = -2\MSusy$) in order to obtain $m_h\ge
114.4$ GeV in the Higgs decoupling limit (this assumes all squark,
slepton, and gaugino soft mass parameters are equal to $\MSusy$,
$\mu=200$ GeV, $m_t = 173$ GeV, and all the soft trilinear couplings
are equal to $A_t$). This gives a value of $\delta\rho$ = 0.0014.
Increasing $m_{\tilde{t}}$ while keeping all other parameters fixed
decreases $\delta\rho$, and for $m_{\tilde{t}}$ = 420 GeV,
$\delta\rho$ is consistent with its 1$\sigma$ upper bound of 0.0009
found in the latest PDG review \cite{PDBook}. With $m_{\tilde{t}}$ =
530 GeV, $\delta\rho$ is consistent with its 1$\sigma$ upper bound
of 0.0006 found in the previous PDG review
\cite{Eidelman:2004wy}.\footnote{As
 this paper was being completed, we noticed that version 2.5.1 of
 $\mathtt{FeynHiggs}$ now uses
 sbottom masses with the SM and MSSM QCD corrections added when
 calculating $\delta\rho$. This
 can give different values of $\delta\rho$, especially for small
 sbottom masses, and it makes $\delta\rho$ more sensitive to $\mu$.
 The results quoted in this paragraph change as follows.  For $m_{\tilde{t}} = 283$ GeV,
 $\delta\rho$ = 0.0011. Increasing $m_{\tilde{t}}$ to 310 GeV gives
 $\delta\rho$ = 0.0009, and $m_{\tilde{t}}$ = 380 GeV gives
 $\delta\rho$ = 0.0006.
 Qualitatively the conclusions
 presented in this section are unaffected. We thank
 S. Heinemeyer for clarifying the difference between the older and
 newer versions.}

The $S$-parameter in the MSSM is in general not very important
\cite{PDBook}.  We estimated it using the formulae in
\cite{Inami:1992rb}.  Including contributions from all squarks and
sleptons, the S-parameter does not reach a value higher than about
0.05 for $\MSusy=\MSusyMin$ in those cases that have large stop
mixing, with the main contribution coming from the stop/sbottom doublet. For
intermediate and vanishing mixing it is negligible. The constraint
on $S$ depends on $T$, but the 1$\sigma$ upper bound on
$S$ is about 0.07 for $T = 0$, whereas a positive value for $T$ allows for
larger values of $S$. Thus the $S$-parameter is a weaker constraint
on the stop masses than the LEP Higgs sector bounds.

\subsection{Constraints from $\bsg$}\label{Sec:bsg}

New physics can contribute at one loop to the decay $\bsg$, and can
therefore be just as important as the SM contribution mediated by a
$W$-boson and the top quark.  This makes the decay $\bsg$ an
important tool in constraining new physics.

The SM contribution to the branching ratio $\mathcal{B}(\bsg)$ is
predicted to be

\beq \label{Eqn:SM bsg} \mathcal{B}(\bsg)_{\textrm{\tiny{SM}}}
\simeq (2.98 \pm 0.26) \times 10^{-4}, \eeq

\noindent see \cite{Becher:2006pu}, whereas the experimental bound
is given by

\beq \label{Eqn:Exp bsg} \mathcal{B}(\bsg)_{\textrm{\tiny{expt}}}
\simeq (3.55 \pm 0.26) \times 10^{-4}, \eeq

\noindent see \cite{:2006bi}.  This allows, but does not require,
new physics contributions \cite{Becher:2006pu}.

There are several contributions to the decay $\bsg$ from the
additional particles in the MSSM, which we now discuss.

Within the Higgs sector, the charged Higgs ($H^{+}$) contributes at
one loop to the decay $\bsg$. The contribution is larger for smaller
$m_{H^{+}}$.  If one only considers this contribution, as one would
in the two-Higgs-doublet model of type II (2HDM (II)), then this
sets a rather stringent lower bound on $m_{H^+}$. The bound of
course depends on the SM prediction and experimental measurement of
$\mathcal{B}(\bsg)$, and in the past used to be about $m_{H^+}
\gtrsim 350$ GeV, see \cite{Gambino:2001ew}, \cite{Hurth:2003vb} and
references therein. The latest results quoted in equations
(\ref{Eqn:SM bsg}) and (\ref{Eqn:Exp bsg}) are expected to change
this bound slightly, but we do not explore this in more detail
\cite{Becher:2006pu}. It is clear, however, that this bound is much
stronger than the bound coming from a direct search of $H^+$ at LEP
which is given by $m_{H^+} \gtrsim 78.6$ GeV \cite{LHWG:2001xy}.
Note that the charged Higgs contribution is mostly independent of
$\tan\beta$; only for very small values of $\tan\beta$ does it
increase substantially.

The charged Higgs, thus, does not contribute much to $\bsg$ in the
decoupling limit for large $m_A$, since here $m_{H^{+}}$ is large.
In the region $m_h \simeq 93$ GeV with $\cossqbma \ge 0.8$ and $m_H
\ge 114.4$ GeV, however, $m_{H^{+}}$ $\simeq$ 125 GeV.  The
contribution from the charged Higgs to $\mathcal{B}(\bsg)$ is then
roughly $7.7 \times 10^{-4}$, more than a factor of two larger than
the SM contribution. We estimated this using version 2.5.1 of the
program $\mathtt{FeynHiggs}$,\footnote{Note that
$\mathtt{FeynHiggs}$ gives $\mathcal{B}(\bsg)_{\textrm{\tiny{SM}}}
\simeq 3.63 \times 10^{-4}$ which is larger than the latest value
quoted in equation (\ref{Eqn:SM bsg}).  This is not of qualitative
importance here.} in the limit of large sparticle masses. Therefore,
the constraint on $\mathcal{B}(\bsg)$ rules out this region of the
Higgs parameter space if one only considers the charged Higgs
contribution.

There are, however, also chargino, neutralino and gluino contributions to
$\bsg$ within the MSSM with minimal flavor violation (MFV).\footnote{There are other
  possibilities for flavor violation within the MSSM, and therefore additional
  contributions to $\bsg$ are possible.  The additional flavor
  violation is small, however, assuming that the only source of flavor
  violation comes from the mixing among the squarks and assuming that
  this is of the same form as the mixing among the quarks,
  i.e.~described by the Cabibbo-Kobayashi-Maskawa (CKM) matrix.  This
  assumption is usually called minimal flavor violation (MFV).
  The MSSM with general flavor violation allows for more contributions to the decay
  $\bsg$, which can sometimes weaken constraints on parameters in the
  MSSM with MFV \cite{Okumura:2004fh}.} NLO contributions can be very
important and need to be included in order to get an accurate
estimate of $\mathcal{B}(\bsg)$ \cite{Hurth:2003vb}. The
contribution from a chargino together with a stop in the loop is
often the most important one.  The chargino-stop contribution can
become very large for small chargino and small stop masses, and it is
proportional to $\tan\beta$ in the amplitude.  However, it vanishes
in the limit of large stop or chargino masses.  From studying the
mSUGRA model, it is known that usually the chargino-stop contribution
to the branching ratio interferes constructively with the SM and the
charged Higgs contribution if the sign of $\mu A_t$ is
positive, whereas it interferes destructively if the sign of $\mu A_t$
is negative \cite{Hurth:2003vb}.

This means that the region $m_h \simeq 93$ GeV is not necessarily
ruled out, since a light stop and a light chargino could cancel the
charged Higgs contribution
\cite{Kim:2006mb,Drees:2005jg,Hooper:2005yg}. Using version 2.5.1 of
$\mathtt{FeynHiggs}$ to calculate the branching ratio of $\bsg$, we
verify this claim in the case of intermediate and larger stop
mixing, at least for $\tan\beta$ not too small.  We find that the
contribution to $\bsg$ from the chargino-stop loop can easily be
large enough to interfere destructively with the charged Higgs
contribution and thus give an experimentally allowed value of
$\mathcal{B}(\bsg)$.  Moreover, in some cases for sizeable stop
mixing, the chargino-stop contribution can be made much larger than
the SM and charged Higgs contribution.  Thus, an experimentally
consistent value of $\mathcal{B}(\bsg)$ can also often be obtained
by finding a chargino mass that gives a chargino-stop amplitude
equal to the negative of the charged Higgs amplitude plus the
negative of twice the SM amplitude. We note that an experimentally
consistent value for $\mathcal{B}(\bsg)$ can always be found without
requiring the stop masses to be larger than $\MSusyMin$, but by
adjusting the chargino mass alone.

If $\tan\beta$ is small enough then $\MSusyMin$ becomes exponentially
large, and the constraint on
$\mathcal{B}(\bsg)$ rules out the $m_h\simeq 93$ GeV region since the chargino-stop
contribution cannot cancel the charged Higgs contribution.

In the case of vanishing stop mixing with \emph{degenerate} stop
soft masses, $\MSusyMin$ is so large that the chargino-stop
contribution to $\bsg$ is too small to cancel the charged Higgs
contribution. However, even in the no-mixing case one of the stops
can be chosen to be light by setting one of the stop soft masses to
a small value. In this case the other stop soft mass needs to be
very large in order for the radiative corrections to the Higgs
sector to be large enough to satisfy the LEP bounds. One light stop,
however, is able to give a sizeable chargino-stop contribution that
can cancel the charged Higgs contribution. For example, we find
$\MSusyMin \simeq 1100$ GeV for $m_h \simeq$ 93 GeV, $\cossqbma\ge
0.8$ and $m_H \ge 114.4$ GeV, with $\tan\beta=20$, $\mu = 200$ GeV,
$m_A\simeq 96$ GeV, and $m_t = 173$ GeV (this assumes that all
squark, slepton, and gaugino soft masses are equal to $\MSusy$, and
all the soft trilinear couplings are equal to $A_t$). Since the
charged Higgs then essentially provides the only contribution to
$\bsg$ beyond that of the SM itself, the branching ratio is again
about $7.7 \times 10^{-4}$.  However, choosing, for example,
$m_{\tilde{t}_L} = 350$ GeV, $m_{\tilde{t}_R} = 2000$ GeV, all the
gaugino soft masses equal to $m_{\tilde{t}_L}$, and keeping all
other squark and slepton soft masses equal to 1100 GeV, gives a
consistent branching ratio of $3.6 \times 10^{-4}$.

\begin{figure}\begin{center}\includegraphics[scale=0.44]{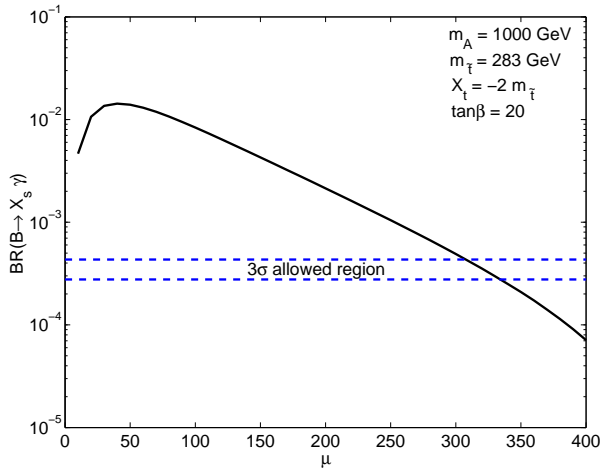}
\caption{$\mathcal{B}(\bsg)$ versus $\mu$ with stop soft masses
$m_{\tilde{t}} \equiv m_{\tilde{t}_L} = m_{\tilde{t}_R} = 283$ GeV,
and natural maximal stop mixing ($X_t/ \MSusy = -2$).  All squark,
slepton, and gaugino soft mass parameters are equal to the stop soft
masses, $m_t = 173$ GeV, $m_A=1000$ GeV, $\tan\beta = 20$, and all
the soft trilinear couplings are equal to $A_t=X_t+\mu\cot\beta$.
\label{Fig:BR_B_to_X_s_gamma_vs_mu_max_mix_TB20}}
\end{center}\end{figure}

In the Higgs decoupling limit, for which $m_h \ge 114.4$ GeV, the
charged Higgs contribution vanishes. Since $\MSusyMin$ is large for
very small $\tan\beta$ or vanishing stop mixing, the chargino-stop
contribution to $\mathcal{B}(\bsg)$ is small, and there is no
inconsistency with the experimental bound. On the other hand,
$\MSusyMin$ can be so low for appreciable amounts of mixing (and if
$\tan\beta$ is not too small) that the chargino-stop contribution
can easily be too large. In this case, however, we can find a
chargino mass that gives a branching ratio of $\bsg$ within the
experimentally allowed region, and again we find no further
constraint on $\MSusy$.  We can achieve this by setting the chargino
mass to a very large value, in which case the chargino-stop
contribution becomes vanishingly small. For negative $\mu A_t$,
however, the chargino-stop loop interferes destructively with the SM
contribution so that we can also adjust the chargino mass until the
chargino-stop amplitude is equal to the negative of twice the SM
amplitude.  This is what happens in the case depicted in
Fig.~\ref{Fig:BR_B_to_X_s_gamma_vs_mu_max_mix_TB20}, where we show
the branching ratio of $\bsg$ as a function of $\mu$. In this
figure, all squark, slepton, and gaugino soft masses are equal to
the stop soft masses, which are given by $\MSusyMin = 283$ GeV, $m_t
= 173$ GeV, $\tan\beta = 20$, $X_t= -2 m_{\tilde{t}}$, and all the
soft trilinear couplings are equal to the stop soft trilinear
coupling, $A_t$.  We find an experimentally allowed value for
$\mathcal{B}(\bsg)$ in this case by choosing $\mu\simeq 330$ GeV. We
note that $\mu$ has to be chosen within about a 30 GeV window for
$\mathcal{B}(\bsg)$ to fall within the 3$\sigma$ allowed region.

\subsection{Constraints from $\Bmumu$}

The decay $\Bmumu$ has not yet been observed. The SM contribution to
this decay is dominated by penguin diagrams involving the $Z$-boson
and box diagrams involving the $W$-bosons \cite{Inami:1980fz}. (The
SM Higgs does contribute to the decay $\Bmumu$ within the SM, but
relative to the dominant contribution it is suppressed by
$m_{\mu}m_{b,s}/m_W^2$, where $m_{\mu}$, $m_b$ and $m_s$ are the
masses of the muon, b-quark and s-quark, respectively, and $m_W$ is
the mass of the $W$-bosons \cite{Grzadkowski:1983yp}.) The SM
contribution to the branching ratio is quite small since it is fourth
order in the weak interactions. It is predicted to be

\beq \mathcal{B}(\Bmumu)_{\textrm{\tiny{SM}}} = (3.42 \pm 0.54)
\times 10^{-9} \eeq

\noindent (see \cite{Abulencia:2005pw} and references therein). This
is well below the current experimental bound from the CDF experiment
at the Tevatron given by

\beq\label{Eqn:Bmumu_CDF_bound}
\mathcal{B}(\Bmumu)_{\textrm{\tiny{expt}}} < 1.5\times 10^{-7} \eeq

\noindent at the 90$\%$ confidence level \cite{Abulencia:2005pw}.

There are several contributions to the decay $\Bmumu$ from the
additional particles in the MSSM, which we now discuss.

The contributions to the decay $\Bmumu$ coming only from the MSSM
Higgs sector are the same as those found in the 2HDM (II). They can
be enhanced by two powers of $\tan\beta$ in the amplitude, which can
compensate for the suppression by the muon mass. One can set an
approximate bound on $m_{H^+}$ assuming this is the \emph{only}
contribution within the MSSM. This bound depends on $\tan\beta$, but
for $\tan\beta=50$ one finds an experimentally allowed value for
$\mathcal{B}(\Bmumu)$ if $m_{H^+}\gtrsim 35$ GeV (see for example
\cite{Logan:2000iv,Chankowski:2000ng}).  As we discussed in Section
\ref{Sec:bsg}, within the 2HDM (II) the constraint on
$\mathcal{B}(\bsg)$ alone forces $m_{H^+}$ to be larger than about
350 GeV. Such a large value for $m_{H^+}$ guarantees that
$\mathcal{B}(\Bmumu)$ is roughly of the same size as the SM result
even for quite large $\tan\beta$, so that it alone provides no
further constraint on the parameter space within the 2HDM (II)
\cite{Bobeth:2001sq}.

In the MSSM there are, however, further contributions to the decay
$\Bmumu$ coming from box and penguin diagrams that involve charginos
and up-type squarks
\cite{Chankowski:2000ng,Bobeth:2001sq,Babu:1999hn,Buras:2002vd,Dedes:2002er,Mizukoshi:2002gs,Dedes:2003kp}.
The penguin diagrams also contain the neutral Goldstone and Higgs
bosons.  The self-energy MSSM Higgs penguin diagrams give the leading
contribution to $\mathcal{B}(\Bmumu)$ for non-negligible mixing in the
stop sector. (In an effective Lagrangian
approach these diagrams may be viewed as inducing a non-holomorphic
coupling between down-type quarks and the up-type Higgs field.) For
large $\tan\beta$, this leading contribution is roughly proportional
to $A_t^2 \tan^6\beta/m_A^4$, and can thus be significantly larger
than the contributions from the Higgs sector alone. Moreover, this
contribution becomes small for very small $\mu$. This contribution does not
vanish for degenerate squark masses, nor in the limit of large
sparticle masses.  Thus, although the branching ratio depends on the
size of the stop masses, it is much more sensitive to the size of
the Higgs masses, $\tan\beta$ and the amount of stop mixing. A light
Higgs sector can give a branching ratio of $\Bmumu$ that
is more than three orders of magnitude above the SM prediction and
thus well ruled out, especially if the stop mixing and $\tan\beta$ are
large.  Moreover, this is the case even for large sparticle masses.
Furthermore, such large values for $\mathcal{B}(\Bmumu)$ can be
reached within the MSSM without violating any other constraints,
including, for example, those on $\mathcal{B}(\bsg)$
\cite{Bobeth:2001sq,Bobeth:2002ch}.

There are further contributions to $\mathcal{B}(\Bmumu)$ which also
have a $\tan^6\beta/m^4_A$ behavior, even assuming that the CKM
matrix is the only source of flavor violation in the squark sector.
These appear if the left-handed up-type soft squark masses of the
three generations are not all equal, so that the left-handed
down-type soft squark mass-squared matrix has off-diagonal terms.
These lead to contributions from loops involving a neutralino or a
gluino and a down-type squark
\cite{Chankowski:2000ng,Babu:1999hn,Mizukoshi:2002gs,Dedes:2003kp,Bobeth:2002ch}.
Cancelations between the chargino and gluino contributions can occur
and the neutralino contribution, although usually smaller, can then
be important (see, for example, \cite{Bobeth:2002ch}).

We estimated the values for $\mathcal{B}(\Bmumu)$ with the program
\href{http://wwwlapp.in2p3.fr/lapth/micromegas/}{$\mathtt{MicrOMEGA}$}
1.3 \cite{Belanger:2001fz,Belanger:2004yn} and the subroutine
\href{http://www.hep.fsu.edu/~isajet/}{$\mathtt{IsaBMM}$} from
$\mathtt{IsaTools}$/$\mathtt{IsaJet}$
\cite{Mizukoshi:2002gs,Paige:2003mg}. We find that the branching
ratio of $\Bmumu$ is well within experimental limits for the region
$m_h$ $\simeq$ $93$ GeV, $m_H\ge114.4$ GeV and $\cossqbma \ge 0.8$
in the case of no (or very little) stop mixing and degenerate squark
soft masses. For intermediate mixing with degenerate squark masses
near $\MSusyMin$, $\mathcal{B}(\Bmumu)$ is consistent with
experimental limits for $\tan\beta \lesssim 20-25$. For natural
maximal mixing, $\mathcal{B}(\Bmumu)$ is consistent with
experimental limits for $\tan\beta \lesssim$ 15.  For larger
$\tan\beta$, as well as for non-degenerate squark soft masses, a
scan over all relevant MSSM parameters is necessary in order to see
whether we can find an experimentally consistent value of
$\mathcal{B}(\Bmumu)$ for such a light Higgs sector. However, for
large stop mixing, it will become increasingly difficult to find a
parameter set that gives a branching ratio consistent with
experiments as $\tan\beta$ is increased. Of course, this assumes
that there are no fortuitous cancelations between the different
contributions, and also that there are no other flavor-violating
contributions such as from R-parity violating couplings. A scan over
the relevant MSSM parameters, even assuming MFV, is beyond the scope
of this paper. The reader is referred to the references found in the
previous two paragraphs, and especially \cite{Hooper:2005yg},
bearing in mind that the current CDF bound on $\mathcal{B}(\Bmumu)$,
equation (\ref{Eqn:Bmumu_CDF_bound}), is stronger than the one used
in these references.

In the Higgs decoupling limit, the dominant flavor-violating effects
involving loops of neutral Higgs bosons decouple, and these large
contributions to $\Bmumu$ become negligible.  Using
$\mathtt{MicrOMEGA}$ 1.3, one may explicitly check that the decay
$\Bmumu$ does not provide stronger constraints on the stop masses
than do the LEP Higgs bounds in the decoupling limit found in
Section \ref{Sec:m_h_greater_114_GeV}.

\section{Implications for Electroweak Symmetry \\ Breaking} \label{Sec:Implications for
EWSB}

In Section \ref{Sec:Lower bounds on MSusy}, we presented lower
bounds on the stop masses consistent with the LEP Higgs bounds, and
in Section \ref{Sec:Constraints from new physics}, we discussed
whether the electroweak $S$- and $T$-parameter and the decays
$B\rightarrow X_s\gamma$ and $B_s\rightarrow \mu^+ \mu^-$ indirectly
put further constraints on the Higgs and/or stop sector. In this
section, we look at the implications for electroweak symmetry
breaking.

The mechanism of radiative electroweak symmetry breaking arises
rather naturally in supersymmetric extensions of the Standard Model
\cite{Ibanez:1982fr,Ellis:1982wr,Alvarez-Gaume:1983gj}. Because of
the large top Yukawa coupling, quantum fluctuations of the stop
squarks significantly modify the up-type Higgs potential, as studied
numerically for the physical Higgs boson mass in the previous
sections. The leading effect, however, is a tachyonic contribution
to the up-type Higgs soft supersymmetry breaking Lagrangian mass.
Over much of parameter space this tachyonic contribution is sufficient
to result in a stop squark quantum fluctuation-induced phase
transition for the Higgs fields, which is generally referred to as
radiative electroweak symmetry breaking.

The leading quantum contribution to the up-type Higgs soft mass
comes from renormalization group evolution below the supersymmetry
breaking messenger scale. The one-loop $\beta$-function for the
up-type Higgs soft mass-squared is, neglecting effects proportional to
gauge couplings,

\beq 16 \pi^2 \beta_{m^2_{H_u}} \simeq 6
  \lambda_t^2 ( m^2_{H_u}+ m_{\tilde{t}_L}^2 + m_{\tilde{t}_R}^2
    + |A_t|^2   )
\label{betamhus} \eq The light Higgs mass bounds require rather
large stop masses and/or stop mixing, where the stop soft trilinear
coupling is related to the mixing parameter by $A_t = X_t + \mu \cot
\beta$. This implies that the stop contributions to the
$\beta$-function in (\ref{betamhus}) proportional to the combination
$(m_{\tilde{t}_L}^2 + m_{\tilde{t}_R}^2 + |A_t|^2)$ are also
sizeable, at least at the low scale. Moreover, for generic
parameters this combination remains sizeable over the entire
renormalization group trajectory up to the messenger scale. For
generic messenger scale values of the up-type Higgs soft mass
squared, $m^2_{H_u}$, the large value of the combination
$(m_{\tilde{t}_L}^2 + m_{\tilde{t}_R}^2 + |A_t|^2)$, along with the
sizeable coefficient in the $\beta$-function (\ref{betamhus}), then
imply that $m^2_{H_u}$ evolves relatively rapidly under
renormalization group evolution.

This evolution is towards tachyonic values of $m^2_{H_u}$ which
reduce the magnitude of the $\beta$-function (\ref{betamhus}). For
running into the deep infrared, the up-type Higgs mass squared would
be driven to values near the zero of the $\beta$-function
(\ref{betamhus}) for which

\beq m^2_{H_u} \simeq - ( m_{\tilde{t}_L}^2 + m_{\tilde{t}_R}^2
    + |A_t|^2).
    \label{zerobeta}
\eq Although this relation is not strictly obtained with finite
running, the up-type Higgs mass squared can approach this value for
very high messenger scale. In Table \ref{Table:combination}, we show
the minimum allowed values of the combination $(m_{\tilde{t}_L}^2 +
m_{\tilde{t}_R}^2 + |A_t|^2)^{1/2}$ deduced from the results of
section \ref{Sec:m_h_greater_114_GeV} consistent with $m_h \geq
114.4$ GeV in the Higgs decoupling limit for large $\tan \beta$. The
minimum allowed values increase with decreasing $\tan \beta$.

\begin{table}\begin{center}\begin{tabular}{|c|c|c|c|}
\hline $X_t/m_{\tilde{t}}$ & $m_t = 168$ GeV & $m_t = 173 $ GeV &
$m_t=178$ GeV \\
\hline 0 & 3630 & 1780 & 1240 \\
\hline -1 & 1460 & 1000 & 770 \\
\hline -2 & 680 & 690 & 710 \\
\hline
\end{tabular}\caption{Minimum allowed values of the combination
$(m_{\tilde{t}_L}^2 + m_{\tilde{t}_R}^2 + |A_t|^2)^{1/2}$ consistent
with a physical Higgs boson mass of $m_h \geq 114.4$ GeV in the
Higgs decoupling limit for large $\tan \beta$, taking into account
only the LEP Higgs sector bounds. The minimum allowed values
increase with decreasing $\tan \beta$.
\label{Table:combination}}\end{center}\end{table}

The full Lagrangian mass squared for the up-type Higgs is a sum of
the soft mass squared and square of the superpotential Higgs mass,
$m^2 = m^2_{H_u} + | \mu|^2$. To leading order in $1/\tan^2 \beta$,
and ignoring the finite quantum corrections to the Higgs potential
which are not of qualitative importance for the present discussion,
this is equal to minus half the $Z$-boson mass squared in the ground
state with broken electroweak symmetry

\beq \frac{1}{2} m_Z^2 \simeq -( m^2_{H_u} + |\mu|^2). \eq

\noindent For $m^2_{H_u}$ near the zero of its $\beta$-function
given by (\ref{zerobeta}), the bounds given in Table
\ref{Table:combination} imply that obtaining the observed value of
the $Z$-boson mass, $m_Z \simeq 91$ GeV, requires a rather sensitive
cancelation between the up-type Higgs soft mass and $\mu$-parameter.
The numerical magnitude of this tuning (which has come to be known
as the supersymmetric little hierarchy problem) is apparent in the
numerical data in Table \ref{Table:combination}, at least for
regions of parameter space which are driven under renormalization
group flow to near the zero of the $\beta$-function
(\ref{betamhus}).

The minimum allowed value of the combination $(m_{\tilde{t}_L}^2 +
m_{\tilde{t}_R}^2 + |A_t|^2)$ for a given lower limit on the Higgs
mass decreases with increasing stop mixing. This may be understood
from the leading expression for the quantum corrected Higgs mass
given in equation (\ref{Eqn:delta_M_uu}). For no stop mixing,
$X_t=0$, the leading correction to the Higgs mass squared comes only
from renormalization group running of the Higgs quartic coupling
below the stop mass scale, and is therefore proportional to $\ln(
m^2_{\tilde{t}} / m_t^2 )$. A linear increase in the Higgs mass
squared in this case requires an exponential increase in $\MSusy$.
However, the stop mixing correction to the Higgs mass squared with
$X_t \neq 0$ comes from a finite threshold correction to the Higgs
quartic coupling at the stop mass scale and is independent of
$\MSusy$ for fixed $X_t / m_{\tilde{t}}$.  In this case a linear
increase in the Higgs mass squared only requires a linear increase
in $(X_t / m_{\tilde{t}})^2$. So increasing stop mixing allows
exponentially lighter stop masses in order to obtain a given Higgs
mass. While such a decrease clearly reduces the soft stop mass
contributions to $\beta_{m^2_{H_u}}$
\cite{Kitano:2006gv,Kane:2004tk} this is partially offset by an
increase in the mixing contribution from the stop trilinear
coupling. From the data in Table \ref{Table:combination}, it is
clear that large stop mixing can decrease the magnitude of
$\beta_{m^2_{H_u}}$ (\ref{betamhus}) by up to a factor of a few
depending on the top mass.  However, the magnitude of the total stop
contribution including mixing is still quite sizeable for a Higgs
mass bound of $m_h \geq 114.4$ GeV.  So large stop mixing alone
cannot appreciably ameliorate the tuning of supersymmetric
electroweak symmetry breaking or satisfactorily solve the
supersymmetric little hierarchy problem.\footnote{Although this
conclusion is valid for a generic choice of messenger scale values
for the sparticle masses, it is possible to reduce the amount of
tuning coming from the running of $m^2_{H_u}$ by a more judicious
choice.  One example is to choose negative stop masses squared at
the high scale which allows the contribution to the tuning from the
running of $m^2_{H_u}$ to be arbitrarily small, as well allow for
the (natural) maximal mixing scenario to be radiatively generated at
the low scale \cite{Dermisek:2006ey}.}

This conclusion essentially remains unchanged for a physical Higgs
boson mass of $m_h \simeq 93$ GeV with $\cossqbma \ge 0.8$ and $m_H
\ge 114.4$ GeV, as seen from the numerical results in Section
\ref{Sec:m_h_simeq_93_GeV}.  In general, one should bear in mind
that indirect constraints on new physics, especially from
$\mathcal{B}(\Bmumu)$, severely restrict the allowed MSSM parameter
space for $m_h\simeq 93$ GeV (see Section \ref{Sec:Constraints from
new physics}).  However, for less than maximal stop mixing, the stop
masses can be somewhat smaller for moderate $\tan\beta$ near the
Higgs non-decoupling limit than in the Higgs decoupling limit (see
also \cite{Kim:2006mb}). The combination $(m_{\tilde{t}_L}^2 +
m_{\tilde{t}_R}^2 + |A_t|^2)^{1/2}$ is in fact the smallest in the
Higgs non-decoupling region near intermediate values for the stop
mixing and for $\tan\beta$ near 10.  It reaches as low as about 650
for $m_t=178$, $\tan\beta = 10.5$, $X_t = -\MSusy$, and gaugino
masses equal to $\MSusy$.  It can be decreased slightly further by
setting the bino and wino masses to smaller values. (For maximal
stop mixing, the combination is actually larger since here the
Tevatron bound on the lighter stop mass forces the stop soft masses
to be larger than required from the LEP Higgs bounds alone.) The
combination always remains sizeable though, and thus the tuning of
electroweak symmetry breaking cannot be ameliorated by much in the
$m_h\simeq 93$ GeV region.

\section{Conclusions}\label{Sec:Conclusion}

The dominant radiative corrections to the tree-level CP-even Higgs
mass matrix, which determines $m_h$ and $m_H$, come from loops
involving the top quark and stop squarks, with larger stop masses
implying larger radiative corrections.  In this paper, we presented
lower bounds on the stop masses consistent with the LEP Higgs bounds
in two different regions in the MSSM Higgs parameter space. The one
region is the Higgs decoupling limit, in which the bound on the mass
of the lighter Higgs is equal to the bound on the SM Higgs, $m_h \ge
114.4$ GeV.  The other region is near the Higgs ``non-decoupling''
limit with $m_h \simeq 93$ GeV in which the Higgs sector is required
to be light. In the latter region, there are two additional
constraints.  One is on the mass of the heavier Higgs, which now
behaves like the SM Higgs, i.e.~$m_H\gtrsim 114.4$ GeV. The other
constraint is on size of the coupling of the lighter Higgs to two
$Z$ bosons which is controlled by the parameter $\sinsqbma$ and here
needs to be less than about 0.2 (i.e.~$\cossqbma \gtrsim 0.8$) for
the lighter Higgs to have escaped detection at LEP.  We denote the
lower bounds on the stop masses consistent with the LEP Higgs bounds
by $\MSusyMin$.

We presented $\MSusyMin$ as a function of $\tan\beta$ in both these
regions in the Higgs parameter space for a variety of MSSM parameter
choices. In particular, we further elucidated the importance of the
top mass and stop mixing, and investigated numerically how larger
top masses and more stop mixing allow for substantially smaller
values of $\MSusyMin$.  We also showed numerically how larger
gaugino masses and larger values of $\mu$ increase $\MSusyMin$.
Moreover, we saw how much $\MSusyMin$ increases if $\mu$ is negative
compared to $\mu$ positive if both $\tan\beta$ and the magnitude of
$\mu$ are large. In the non-decoupling region, we discussed how the
constraints on $\cossqbma$ and on $m_H$ lead to increased values for
$\MSusyMin$.

We also considered how $\MSusyMin$ changes as a function of $m_h$.
Since $A_t$ and $M_3$ most naturally have the opposite sign at low
scales due to renormalization group running, a negative value of
$A_t$ is more natural in a convention where $M_3$ is positive. For
negative $A_t$ and stop masses less than a few TeV, the upper bound
of $m_h$ in the MSSM is around 131 GeV.

We demonstrated that the two regions in the Higgs parameter space
have roughly the same $\MSusyMin$ if $\tan\beta$ is large. For
moderate values of $\tan\beta$ and non-maximal stop mixing,
$\MSusyMin$ is larger in the Higgs decoupling region than in the
Higgs non-decoupling region. As $\tan\beta$ decreases, however,
$\MSusyMin$ is larger in the Higgs non-decoupling region than in the
Higgs decoupling region.

We also considered additional constraints from the electroweak $S$-
and $T$-parameter and the decays $B \to X_s \gamma$ and $B_s \to
\mu^+ \mu^-$, which also constrain the Higgs and/or stop sector.

The main contribution to the $T$-parameter within the MSSM usually
comes from the stop/sbottom doublet and, for a given set of
parameters, is larger for larger stop (and sbottom) mixing as well
as for smaller stop and sbottom masses.  We found that the value of
the $T$-parameter is well within its 2$\sigma$ bound for stop masses
equal to $\MSusyMin$. In fact, only for maximal stop mixing do we
find small enough values for $\MSusyMin$ that give a contribution to
the $T$-parameter that does not also fall within its 1$\sigma$
bound. For such large stop mixing one must then increase the stop
masses by a small amount above $\MSusyMin$ to also satisfy the
1$\sigma$ bound on the $T$-parameter.

We found that the contribution to the $S$-parameter is not
large, and that the $S$-parameter therefore does not provide an
additional constraint on the stop masses.

The indirect constraint on $\mathcal{B}(\bsg)$ in many cases does
not provide an additional constraint on the stop masses. In the
Higgs non-decoupling region for $m_h\simeq 93$ GeV, the Higgs sector
is required to be light, and the charged Higgs contribution to
$\bsg$ is large.  The charged Higgs contribution can usually be
canceled by the chargino-stop contribution through a judicious
choice of the chargino mass.  However, for vanishing stop mixing and
assuming degenerate stop soft masses, $\MSusyMin$ is so large that
the chargino-stop contribution is too small to cancel the charged
Higgs contribution. For vanishing stop mixing, we therefore require
non-degenerate stop soft masses with one light stop so that the
chargino-stop contribution can be large enough to give an
experimentally consistent value for $\mathcal{B}(\bsg)$ (the other
stop must then be very heavy so that the LEP Higgs constraints are
satisfied). In the Higgs decoupling limit, the charged Higgs
contribution vanishes. We find no further constraint on the stop
masses.  Even for large stop mixing, for which $\MSusyMin$ can be
very small, one can always obtain an experimentally consistent value
for $\mathcal{B}(\bsg)$ by adjusting the chargino mass.

The main contributions to the flavor-violating decay $\Bmumu$ come
from flavor violating Higgs couplings, and these decouple in the
Higgs decoupling limit.  Thus, the indirect constraint on
$\mathcal{B}(\Bmumu)$ is only important in the Higgs non-decoupling
region. In this region, however, it is able to severely restrict the
allowed parameter space, since the flavor violation does not
decouple in the limit of large sparticle masses. In fact, the region
for such a light Higgs sector is ruled out if stop mixing and
$\tan\beta$ are large, unless there are fortuitous cancelations
amongst the various contributions, or there are additional
flavor-violating contributions from, for example, R-parity violating
couplings that cancel these contributions.

We note that we did not consider the constraint on the anomalous
magnetic moment of the muon, $(g-2)_{\mu}$, since it decouples in
the limit of large sneutrino and smuon masses.  It alone is thus
unable to directly provide a further constraint on the Higgs sector
or on the stop masses.

Lastly, we discussed the implications of our numerical analysis for
electroweak symmetry breaking.  Large stop mixing generically
decreases the tuning of supersymmetric electroweak symmetry
breaking, but is unable to do so sufficiently to solve the
supersymmetric little hierarchy problem. Moreover, the tuning can be
ameliorated only slightly in the $m_h\simeq 93$ GeV region compared
to the $m_h \ge 114.4$ GeV region (for intermediate values of the
stop mixing and moderate values of $\tan\beta$), and thus the
supersymmetric little hierarchy problem cannot be satisfactorily
solved in either of the two regions.

\newpage

\noindent {{\bf \large Acknowledgements}}

\vspace{0.1cm}

I would like to thank my advisor S.~Thomas for suggesting this
project and for his support and guidance throughout. I would also
like to thank S.~Heinemeyer for helpful email exchanges about
$\mathtt{FeynHiggs}$.  It is a pleasure to further acknowledge
useful conversations with J.F.~Fortin, D.E.~Kaplan, A.~Lath,
N.~Sehgal, and N.~Weiner, as well as R.~Derm{\'i}{\v s}ek who
pointed out two relevant references. I also want to thank J.~Shelton
for reading the manuscript and providing many helpful suggestions.
This research is supported by a Graduate Assistantship from the
Department of Physics and Astronomy at Rutgers University.

\vspace{0.5cm}

\appendix

\section{Mixing in the Two Doublet Higgs Sector} \label{Sec:MSSM Higgs sector and LEP}

A Higgs sector with electroweak symmetry broken to
electromagnetism,\\ $SU(2)_L \times U(1)_Y \rightarrow U(1)_Q$, by
two $SU(2)_L$ doublets, $H_u$ and $H_d$, with hypercharge $Y=\pm 1$
respectively, has two physical scalars, $h$ and $H$, a pseudoscalar,
$A$, and a charged scalar, $H^{\pm}$. The couplings of the scalar
mass eigenstates, $h$ and $H$, to the gauge bosons are determined by
the associated amplitudes of the neutral components of the gauge
eigenstate doublets, $H_u^0$ and $H_d^0$. It is instructive to
consider various vectors in the ${\rm Re}(H_d^0) - {\rm Re}(H_u^0)$
plane in order to describe these couplings and the relationship
between the mass and gauge interaction eigenstates.

\begin{figure}[!htb]\begin{center}\includegraphics{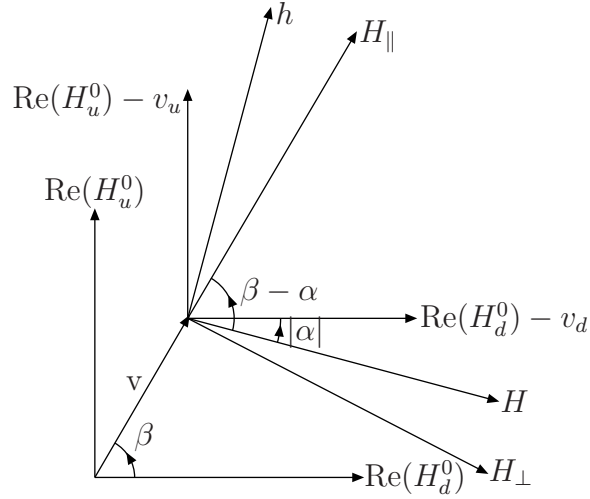}
\caption{Relationship between the ${\rm Re}(H_u)-{\rm Re}(H_d)$ and
$H_{\|}-H_{\bot}$ bases and $h-H$ mass eigenstates for the two
doublet Higgs sector.\label{Fig:bases}}
\end{center}
\end{figure}

Electroweak symmetry is broken by the expectation values
$\langle{\rm Re} (H_u^0)\rangle=v_u$ and $\langle{\rm Re}
(H_d^0)\rangle=v_d$. These expectation values define a vector in the
${\rm Re}(H_d^0) - {\rm Re}(H_u^0)$ plane with an angle $\beta$
defined by $\tan \beta = v_u/v_d$ as indicated in
Fig.~\ref{Fig:bases}. The two physical neutral CP-even scalar mass
eigenstates are fluctuations about the expectation value in this
plane and are related by a rotation to the gauge eigenstates
conventionally defined by an angle $\alpha$ as \cite{Gunion:1989we}

\begin{equation}
\left( \begin{array}{c} H \\ h
\end{array} \right) = \sqrt{2}
\left( \begin{array}{cc} \cos\alpha & \sin\alpha \\
-\sin\alpha & \cos\alpha \end{array} \right)    \left(
\begin{array}{c} {\rm Re}(H_d^0) - v_d \\ {\rm Re}(H_u^0) - v_u \end{array} \right)
\end{equation}

\noindent Vectors in the ${\rm Re}(H_d^0) - {\rm Re}(H_u^0)$ plane
which are parallel and perpendicular, $H_{\|}$ and $H_{\bot}$, to
the expectation value vector may also be defined as indicated in
Fig.~\ref{Fig:bases}

\begin{equation}\label{Eqn:HperpHpar_defn}
\left( \begin{array}{c} H_{\bot} \\ H_{\|} \end{array} \right) =
\sqrt{2} \left( \begin{array}{cc} \sin\beta & -\cos\beta \\
\cos\beta & \sin\beta \end{array} \right)    \left(
\begin{array}{c} {\rm Re}(H_d^0) - v_d \\ {\rm Re}(H_u^0) - v_u \end{array}
\right),
\end{equation}

\noindent see also \cite{Kim:2006mb}.  The physical mass eigenstates
are related to these by a rotation
\begin{equation}\label{Eqn:HperpHpar}
\left( \begin{array}{c} H \\ h \end{array} \right) =
\left( \begin{array}{cc} \sin(\beta - \alpha) & \cos(\beta - \alpha) \\
-\cos(\beta-\alpha) & \sin(\beta-\alpha) \end{array} \right) \left(
\begin{array}{c} H_{\bot} \\ H_{\|} \end{array} \right).
\end{equation}

\noindent The neutral Goldstone pseudoscalar boson, $G$, which is
eaten by the $Z$-boson is by definition the imaginary part of the
linear combination of the components of the neutral Higgs doublets
which are aligned with the expectation value, and the physical
pseudoscalar Higgs boson, $A$, is the perpendicular combination

\begin{equation}\label{Eqn:A and G definition}
\begin{array}{c} A = {\rm Im}(H_{\bot}) \\ G = {\rm Im}(H_{\|})
\end{array}.
\end{equation}

\noindent These states are related to the gauge eigenstates through
a rotation by the angle $\beta$

\beq  \left( \begin{array}{c} A \\ G
\end{array} \right) = \sqrt{2}
\left( \begin{array}{cc} -\sin\beta & \cos\beta \\
\cos\beta & \sin\beta \end{array} \right)    \left(
\begin{array}{c} {\rm Im}(H_d^0) \\ {\rm Im}(H_u^0) \end{array}
\right).
\eq

\noindent The charged Goldstone bosons, $G^{\pm}$, and the charged
Higgs mass eigenstates, $H^{\pm}$, are defined similarly as

\begin{equation}
 \begin{array}{c} H^{\pm} = {\rm Im}(H_{\bot}^{\pm}) \\
 G^{\pm} =   {\rm Im}(H_{\|}^{\pm})   \end{array},
 \label{Hpmdef}
\end{equation}
where $H_{\bot}^{\pm}$ and $H_{\bot}^{\pm}$ are defined in analogy
with equation (\ref{Eqn:HperpHpar_defn}).

We may consider Higgs decoupling limits of the two doublet Higgs
sector in which $m_H, m_A, m_{H^\pm} \gg m_h$ so that only a
single light Higgs doublet remains in the low energy theory. A
particular decoupling limit is one for which the physical mass
eigenstate of the light Higgs doublet is aligned with the
expectation value vector so that $H_{\|}$ is the single Higgs
doublet of the low energy theory and $H_{\bot}$ contains the heavy
mass eigenstates. This is the unique decoupling limit available to
the the tree-level Higgs potential of the MSSM, although other
misaligned decoupling limits may be realized for more general two
doublet potentials. In the aligned decoupling limit $h=H_{\|}$ with
$\sin( \beta- \alpha)= 1$ and $\cos( \beta- \alpha)= 0$.

The couplings of physical Higgs bosons to gauge bosons arise from
the gauge kinetic terms of the Higgs fields

\beq (D_{\mu} H_u)^* D^{\mu} H_u + (D_{\mu} H_d)^* D^{\mu} H_d
\label{higgskinetic}
\eq
where $D_{\mu}= \partial_{\mu} + i
g^{\prime} \frac{1}{2} Y B_{\mu} + i g T^a W^a_{\mu}$ is the
covariant derivative including the $SU(2)_L \times U(1)_Y$ gauge
connections $W_{\mu}^a$ and $B_{\mu}$. A coupling of two gauge
bosons to a single physical Higgs boson arises from
(\ref{higgskinetic}) with a gauge field in each covariant
derivative, a physical Higgs boson in one Higgs field, and an
expectation value in the other Higgs field. In terms of the $H_{\|}
- H_{\bot}$ basis these couplings are particularly simple. Since it
is only $H_{\|}$ which is parallel to the expectation value, only
this component appears in these couplings

\beq (D_{\mu} v)^* D^{\mu} H_{\|} + (D_{\mu} H_{\|})^* D^{\mu} v
\label{dvdh}, \eq
where of course $D_{\mu} v$ contains only gauge
field couplings since $\partial_{\mu} v =0$. In terms of the
physical gauge bosons, the couplings in (\ref{dvdh}) give rise to
$WWH_{\|}$ and $ZZH_{\|}$ interactions. In terms of the physical
Higgs scalar eigenstates $h$ and $H$ related to $H_{\|}$ in
(\ref{Eqn:HperpHpar}) these couplings give interactions $WWh$ and
$ZZh$ proportional to $\sin( \beta- \alpha)$ and interactions $WWH$
and $ZZH$ proportional to $\cos( \beta- \alpha)$. In the Higgs
aligned decoupling limit the latter interactions vanish since
$H=H_{\bot}$ in this limit with $\cos( \beta- \alpha)= 0$. Note that
there are no interactions of two gauge bosons with a single charged
Higgs boson of the form $W^{\pm}ZH^{\mp}$, since from (\ref{Hpmdef})
the physical charged Higgs boson resides in $H_{\bot}$, while from
(\ref{dvdh}) these type of interactions arise only for $H_{\|}$.
This result generalizes to any number of Higgs doublets.

A coupling of a single gauge boson to two physical Higgs bosons
arises from (\ref{higgskinetic}) with a single gauge field in one of
the covariant derivatives, physical Higgs bosons in each Higgs
field, and a derivative acting on one of the Higgs fields

\beq ( D_{\mu} H^*) \partial^{\mu} H + ( \partial_{\mu} H)^* D^{\mu}
H \label{hdhequation}
\eq

\noindent where the covariant derivatives $D_{\mu}$ are again
understood to only contain gauge fields here. This subset of couplings
represents the Higgs current coupling to a single gauge boson, and
therefore must contain at least one imaginary component of a Higgs
field. Now from equations (\ref{Eqn:A and G definition}) and
(\ref{Hpmdef}) the imaginary components of the Higgs fields appear
in the physical mass eigenstates only through $H_{\bot}$. So the
couplings (\ref{hdhequation}) to physical mass eigenstates are
contained in

\beq ( D_{\mu} H_{\bot})^*
\partial^{\mu} H_{\bot} + ( \partial_{\mu} H_{\bot})^* D^{\mu}
H_{\bot}
\eq

\noindent In terms of the physical $Z$ gauge boson these couplings
give rise to $ZH_{\bot}H_{\bot}$ interactions. In terms of the
physical eigenstates $h$ and $H$ related to $H_{\bot}$ in
(\ref{Eqn:HperpHpar}), these couplings give the interaction $ZAH$
proportional to $\sin(\beta- \alpha)$ and $ZAh$ proportional to
$\cos (\beta- \alpha)$. In the Higgs aligned decoupling limit the
latter interaction vanishes since $h=H_{\|}$ in this limit with
$\cos( \beta- \alpha)= 0$.

\section{Quasi-Fixed Point for the Stop Trilinear \\ Coupling $A_t$}\label{Sec:At fixed point}

The MSSM has a number of quasi-fixed points for various couplings
that make a relation in the low energy theory between them and other
parameters quite natural. These couplings include the top Yukawa and
top trilinear coupling. Consider first the so called Pendleton-Ross
quasi-fixed point for the top Yukawa \cite{Pendleton:1980as}. The
one-loop $\beta$-functions for the top Yukawa $\lambda_t$ and
$SU(3)_C$ gauge coupling $g_3$ in the MSSM are

\beq \label{Eqn:RGE_topYukawa} 16 \pi^2 \beta_{\lambda_t} =
\lambda_t \big( 6\lambda_t^2 - \frac{16}{3}g_3^2 \big) \eeq \beq
\label{Eqn:RGE_g3} 16 \pi^2 \beta_{g_3}= -3~g_3^3 \eeq where
$SU(2)_L$ and $U(1)_Y$ gauge interactions have been neglected in
$\beta_{\lambda_t}$. These $\beta$-functions give a one-loop
$\beta$-function for the logarithm of the ratio of couplings of \beq
16 \pi^2 \beta_{ \ln(\lambda_t / g_3)} =  6 \lambda_t^2 -
\frac{7}{3}g_3^2. \eeq Vanishing of this $\beta$-function implies
that the ratio of the top Yukawa to $SU(3)_C$ gauge coupling,
$\lambda_t / g_3$, is independent of renormalization group scale at
one-loop. Since $\beta_{g_3}$ does not vanish at one-loop, $g_3$ is
renormalization scale dependent. So the vanishing of $\beta_{
\ln(\lambda_t / g_3)}$ defines a quasi-fixed point for $\lambda_t$
rather than a scale-independent fixed-point relation. With the above
approximations the Pendleton-Ross quasi-fixed point in the MSSM
occurs for

\beq \lambda_t^2 = \frac{7}{18} g_3^2. \label{PRfp} \eeq

\noindent Since $\beta_{g_3}$ is independent of $\lambda_t$ at
one-loop, and the coefficient of the $\lambda_t^2$ term in $\beta_{
\ln(\lambda_t / g_3)}$ is positive, this quasi-fixed point is
attractive for $\lambda_t/g_3$ both above and below the quasi-fixed
point value. Moreover, since $\beta_{\lambda_t}$ is cubic in $\lambda_t$,
it is very strongly attractive from above.

The top trilinear coupling and gluino mass have a similar
quasi-fixed point relation \cite{Ferreira:1995sn,Lanzagorta:1995ai}.
The one-loop $\beta$-functions for the top trilinear coupling,
$A_t$, and gluino mass, $M_3$,  are

\beq \label{Eqn:RGE_A-term} 16 \pi^2 \beta_{A_t}= 12 \lambda_t^2 A_t
+ \frac{32}{3} g_3^2 M_3 \eeq \beq \label{Eqn:RGE_gluino} 16 \pi^2
\beta_{M_3} = - 2g_3^2 M_3 \eeq where $SU(2)_L$ and $U(1)_Y$ gauge
interactions have been neglected in $\beta_{A_t}$. Adding these
$\beta$-functions gives \beq 16 \pi^2 \beta_{(A_t + M_3)} =
12\lambda_t^2 A_t + \frac{14}{3} g_3^2 M_3. \eeq At the
Pendleton-Ross quasi-fixed point (\ref{PRfp}) for the top Yukawa in
the MSSM this reduces to \beq 16 \pi^2 \beta_{(A_t + M_3)} = {14
\over 3} g_3^2 \left( A_t + M_3 \right). \eeq The vanishing of
$\beta_{ A_t + M_3}$ again defines a quasi-fixed point for $A_t$
rather than a scale independent fixed point relation. With the above
approximations at the Pendleton-Ross quasi-fixed point, the top
trilinear then has a quasi-fixed point of \beq A_t = -M_3.
\label{Atfp} \eeq Since the coefficient of $\beta_{A_t+M_3}$ is
positive, this quasi-fixed point is attractive. Moreover, since it
is proportional to $g_3^2$ with a sizeable coefficient it is rather
strongly attractive. Because of this it is most natural for $A_t$
and $M_3$ to have opposite sign and be comparable in magnitude at
low scales due to renormalization group evolution. This conclusion
is rather insensitive to messenger scale boundary conditions for
$A_t$, at least for large enough messenger scales.

\vspace{0.5cm}

\bibliographystyle{utcaps}
\bibliography{Bibliography}

\end{document}